\documentclass[iop,numberedappendix]{emulateapj}
\usepackage{natbib}  
\usepackage{graphicx}
\usepackage{mathtools}
\usepackage{booktabs}
\usepackage{paralist}
\usepackage{longtable}
\usepackage{threeparttable}
\DeclareGraphicsExtensions{.pdf,.png,.jpg}
\newcommand{\ssmd}{{$\Sigma_{\star}$}}
\newcommand{\str}{{\textit{Stripe82}}}
\newcommand{\sbdim}{{mag arcsec$^{-2}$}}
\newcommand{\ssmddim}{{$M_{\odot}$ pc$^{-2}$}}
\newcommand{\sbeff}{{$\langle \mu \rangle_e$}}
\newcommand{\mtol}{{$M/L$}}

\begin{document}

\title{Deep Surface Brightness Profiles\\ of Spiral Galaxies from SDSS Stripe82:\\ Touching Stellar Halos}

\author{Judit Bakos \& Ignacio Trujillo}

\affiliation{Instituto de Astrof\'isica de Canarias, C/ V\'ia Lactea S/N, La Laguna, Tenerife, Spain }
\affiliation{Departamento de Astrof\'isica, Universidad de La Laguna, E-38205, La Laguna, Tenerife, Spain}

\email{jbakos@iac.es, trujillo@iac.es}

\begin{abstract}

Using SDSS \textit{Stripe82} data we have obtained deep radial surface brightness profiles of 7 face-on to intermediate inclined late-type spirals down to $\mu_{r'} \sim$ 30 mag arcsec$^{-2}$. We do not find any evidence for a sharp cut-off of the light distribution of the disks but a smooth continuation into the stellar halos of galaxies. Stellar halos start to affect the surface brightness profiles of the galaxies at $\mu_{r'} \sim $28 mag arcsec$^{-2}$, and at a radial distance of $\gtrsim 4-10$ inner scale-lengths. We find that the light contribution from the stellar halo could be responsible of previous classification of surface brightness profiles as Type~III in late-type galaxies. In order to estimate the contribution of the stellar halo light to the total galaxy light, we carried out a Bulge/Disk/Stellar Halo decomposition by simoultaneously fitting all components. The light contribution of the halo to the total galaxy light varies from $\sim$ 1$\%$ to $\sim$ 5$\%$, but in case of ongoing mergers, the halo light fraction can be as high as $\sim$ 10$\%$, independently of the luminosities of the galaxies. We have also explored the integrated $(g'-r')$ color of the stellar halo of our galaxies. We find $(g'-r')$ colors ranging from $\sim$ 0.4 to $\sim$ 1.2. By confronting these colors with model predictions, we encounter problems to fit our very red colors onto stellar population grids with conventional IMFs. Very red halo colors can be attributed to stellar populations dominated by very low mass stars of low to intermediate metallicity produced by bottom-heavy IMFs.

\end{abstract}


\vspace{2pc}
\maketitle

\defcitealias{BTP08}{BTP08}
\defcitealias{PT06}{PT06}

\section{Introduction}
\label{sec:intro}

The faint outer disk and stellar halo of spiral galaxies hold a plethora of information related to their formation and evolution. This is because, due to their relatively long star formation and dynamical time-scales, the fossil records left behind by the formation processes survive longer than in the inner parts of the galaxies. Consequently, the characteristics of these faint subcomponents must be closely connected to the evolutionary paths the galaxies might have taken. Studying the structure and stellar populations of the faint outer disks and stellar halos makes possible to test the predictions of the current cosmological paradigm in detail.

The disks of spiral galaxies have been studied in great detail. Structural studies of nearby spiral galaxies often concern the inner, bright structure, the bulge, bar(s), rings, etc \citep[e.g.][]{deJong1996,Graham2001,MacArthur2003}. During the past decade, however, our knowledge of the structure and stellar populations of the disks at faint surface brightness levels has grown abundantly. These studies pushed the surface brightness limit to $\mu_{r'} \sim$ 27 \sbdim being able to explore the faint outskirts of disks. Thanks to these efforts, not only we got to know that a large amount of stars is present (\citealp{Pohlen2002,Erwin2005}; \citealp[][PT06 herafter]{PT06}; \citealp{Erwin2008}) in these regions where star formation is negligible, but they also showed that the paradigm of 'disk-as-exponential' \citep{Patterson1940,deV1958,Freeman1970} is far from accurate. 

Based on their behavior galaxy surface brightness profiles are classified into three major classes termed as Type~I, Type~II, and Type~III. Type~I galaxies follow pure exponential profiles. The Type~II morphology represents all galaxies with a ``downbending break'': a double exponential profile with a steeper outer exponential (\citealp{Pohlen2002}; \citetalias{PT06}; \citealp{Erwin2008}). This class is the revisited and extended version of the Freeman Type~II class \citep{Freeman1970} now including the so-called truncations discovered by \citet{vanderKruit79}. Type~III galaxies show an ``upbending break'', a double exponential profile with a less inclined outer exponential discovered by \citet{Erwin2005}. \citetalias{PT06} showed that only a minority of the surface brightness profiles of late-type disks follow a single exponential decline. The majority of late-type disks, $\sim$ 90$\%$, exhibits so called breaks in the faint outer regions. Apart from the fact that these features can be linked to different morphological features (such as bars, rings, spiral arms, etc) inside the disk, the properties of breaks follow tight correlations with the host galaxy properties, like Hubble-type, total luminosity or stellar mass, and disk scale-length. Type~II phenomenon is not limited to the local universe, but has also been found in galaxies at higher redshift \citep{Perez2004,Trujillo2005}. Moreover, \citet{Azzollini2008a} showed that the break radius could be used as a tracer of size evolution of disk galaxies. In the last $\sim$ 8 Gyrs, Type~II disks went through significant (size) evolution.

The stellar population properties obtained via color profiles revealed further clues and can be used to constrain models \citep{Roskar2008,Sanchez-Blazquez2009,Martinez-Serrano2009} that besides the effect of the motion of the stars (e.g. radial migration) consider the effects of star formation, and/or the threshold in star formation \citep{Kennicutt1989}. \citet{BTP08} showed that these three surface brightness types directly translate into distinct behavior on their (g'-r') color profiles. In Type~II galaxies the color profile follows a prominent ``U-shape'', with the bluest color at the break radius. This suggests that in Type~II galaxies the origin of the break phenomena is related to star formation thresholds \citep[e.g.][]{Schaye2004}. By having extended the analysis to intermediate redshifts, it is seen that the ``U-shape'' of the (g'-r') color profiles is present in the last $\sim$8 Gyrs \citep{Azzollini2008}, revealing that not just the morphology but the \textit{distribution of stellar population content} is a common feature during the evolution of Type~II galaxies. 

The surface brightness levels that are requiered to explore the distribution of stars inside the disks are relatively bright ($\mu_{r'} \lesssim $ 27 \sbdim), making it possible to carry out photometric studies of disks to large distances (even to intermediate redshifts) on statistically representative samples. The stellar halos of spiral galaxies are, however, so faint that to observe them it is necessary to go $\sim$ 10 magnitudes below the night sky level (to reach $\sim$ 30 \sbdim). This requires nothing less but excellent observing conditions free of observational artifacts such as flat-fielding problems or large skysubtraction uncertainties. Hitherto only a few \textit{integrated} photometric studies can be found in the literature that have attempted to reach the surface brightness of stellar halos \citep[e.g.][and references therein]{Zibetti2004,Jablonka2010}. To investigate the statistical properties of stellar halos \citet{Zibetti2004a} and \citet{Bergvall2010} stacked $\sim$ 1000 galaxies arriving at the level of $\mu_{r'} \sim $31 \sbdim. These observations have also revealed anomalous red colors in stellar halos (otherwise known as the ``red halo phenomenon'') that could indicate the presence of extreme stellar populations \citep[see also e.g.][]{Lequeux1998,Zibetti2004a,Zackrisson2006,Zackrisson2012}.

Another way to explore the stellar population properties of stellar halos is achieved by 'resolved star' technique. By probing, for instance, RGB stars inside the stellar halos \citep[e.g.][and references therein]{Ferguson2002,Davidge2003,Ferguson2007,Rejkuba2009,Barker2012}, this method proved to be efficient in reaching surface brightness levels that are $\gtrsim$ 2 magnitude deeper than what was so far accomplished by integrated photometry. However, the number of systems that can be studied by this technique is largely limited by the resolving power of current telescopes that only allows us to resolve stars in galaxies in our close vicinity ($\sim$ 15 Mpc). Despite of this, these studies provided stellar population and spatial contraints of stellar halos with unprecedented certainty. With the current cosmological paradigm in accordance, there is a growing consensus that the stellar halos of nearby galaxies such as M31, NGC~2403, NGC~0891 or M81 are dominated by old but metal-poor populations. There is also compelling evidence that stellar halos are full of tidal debris, remnants of recent mergers with satellites of these giant galaxies. One of the most well studied galaxies, our neighbor, M31 went through quite a exciting merger history and hence is full of substructures \citep[e.g.][]{Ferguson2002,Ibata2007,McConnachie2009,Tanaka2010}, similarly to the stellar halo of the Milky Way \citep[e.g.][]{Ibata1994,Majewski1999,Newberg2002,Gilmore2002,Belokurov2007,Juric2008,Bell2008}. There is also plenty of observational evidence showing ongoing merger phenomenology in spiral galaxies at these surface brightness levels, stellar streams have been detected around disk galaxies like NGC~5907, NGC~4013 \citep{Martinez-Delgado2008,Martinez-Delgado2009} and NGC~0891 \citep{Mouhcine2010}. 

One of the expectations of the current cosmological paradigm is that properties of the stellar halo (e.g. metallicity) correlates with the properties of the host galaxy \citep[e.g.][]{Brook2004,Bullock2005,Cooper2010}. Although, these correlations might be obscured due to stochastic variations in the merger/accretion history. Disks are expected to be less changed by external phenomena but more driven by \textit{secular} processes. To be able to distinguish between the consequences of these two entirely different effects, and to be able to discern the dominant mechanisms that shape spiral galaxies, we need to study the disks and stellar halos simoultaneously in as many systems as possible by reaching low surface brightness levels.

The aim of this paper is to study the structure and stellar content of 7 late-type, intermediate inclined to face-on spiral galaxies down to a surface brightness level of $\mu_{r'} \sim$ 30 \sbdim\ using \textit{integrated light} photometry that beforehand could only be reached by resolved star technique on individual galaxies. With this work we would like to demonstrate that with a \textit{correct data reduction} it is possible to study faint systems like the stellar halo around external galaxies as far as $\sim$ 100 Mpc using the \str\ dataset. This opens the possibility to explore a large variety of galaxy types and masses that are not present in our vicinity.

\begin{table*}[t]
\begin{center}

\caption{\label{tab:sample} The Sample}
\begin{tabular*}{\textwidth}{@{\extracolsep{\fill}} ccccccc }
\toprule
\textbf{Galaxy} & \textbf{$\alpha$} & \textbf{$\delta$}  & \textbf{Hubble} & \textbf{T} & \textbf{D} & \textbf{M$_{abs}$}\\
 & \multicolumn{2}{c}{\textbf{(J2000.0)}}  & type & & \textbf{[Mpc]} & \textbf{[\textit{B}-mag}]  \\
(1) & (2) & (3) & (4) & (5) & (6) & (7)  \\

\hline
NGC0450 & 01 15 30.4 & -00 51 40 & SABc & 5.8$\pm$0.5 & 24.4 & -19.78 \\

NGC0941 & 02 28 27.8 &  -01 09 06 & SABc & 5.3$\pm$0.7 & 21.9 & -19.13 \\

NGC1068 & 02 42 40.7 & -00 00 48 & Sb & 3.0$\pm$0.3 & 15.3 &  -21.50 \\

NGC1087 & 02 46 25.1  & -00 29 55 & SABc & 5.2$\pm$0.8 & 20.7 & -20.65 \\

NGC7716 & 23 36 31.4  & +00 17 50 & SBb & 3$\pm$0.5 & 36.5 & -20.31 \\

UGC02081 & 02 36 00.9  & +00 25 13 & SABc & 5.8$\pm$0.5 &36.5 & -18.53 \\

UGC02311 & 02 49 27.9  & -00 52 23 & SBb & 3.0$\pm$0.4 & 102.8 & -21.50 \\
\bottomrule
\end{tabular*}
\end{center}
\end{table*}

\section{The Sample}

We have created a sample of late-type spiral galaxies selected from the Hyperleda\footnote{http://leda.univ-lyon1.fr/} online catalog applying several criteria for the Hubble-type (T parameter), the axis ratio (major axis/minor axis), the absolute B-magnitude (M$_{abs}$), and the radial velocity relative to the Local Group (corrected for virgocentric inflow, \textit{v}$_{vir}$).

We restricted the Hubble-type to cover the types between Sbc and Sdm (2.99 $<$ T $<$ 8.49), building up an intermediate- to late-type sample. The axis ratio is selected to be log\textit{r}$_{25}$ $<$ 0.301 (corresponding to an inclination of $\leq$ 61$^{\circ}$). This allows us to have a face-on to intermediate inclined galaxy sample, and to mitigate problems related to dust extinction. The distance of our galaxies was restricted accordingly to create a volume-limited but complete sample of galaxies with M$_{abs}$ $<$ -18.4 B-mag (see \citetalias{PT06}). To this end, we selected galaxies that conform a radial velocity (corrected for the Virgo infall) \textit{v}$_{vir}$ $<$ 3250 km sec$^{-1}$ (out to $\sim$ 46 Mpc estimated following the Hubble relation with H$_0$ = 70 km s$^{-1}$ Mpc$^{-1}$). Hyperleda yields a sample $\sim$ 800 galaxies with no coordinate restrictions applied. To identify the galaxies that fall within the area ($\sim$ 270 square-degrees) of \str, we carried out a cross correlation between the full Hyperleda sample and the \str\ catalog. We also added a luminosity extended sample into this cross correlation, containing late-type spirals brighter than M$_{abs}$ $<$ -20.5 B-mag by allowing \textit{v}$_{vir}$ $<$ 7000 km sec$^{-1}$. 

In total in this paper we explored the following 7 spiral galaxies:  NGC~0450, NGC~0941, NGC~1068, NGC~1087, NGC~7716, UGC~02081, and UGC~02311. The sample properties can be found in Table \ref{tab:sample}.

\section{Data and Profile Extraction}

\subsection{The Data}

Our imaging data come from the Sloan Digital Sky Survey \str. This stripe covers about $\sim 270$ square degrees of the sky in the equatorial plane and has been observed in all SDSS bands multiple, often more than 80, times as part of the SDSS Supernovae Legacy Survey \citep{Abazajian2009}. Even though this particular survey was carried out to study variable objects, this implies that after stacking the data is $\sim$ 2 magnitudes deeper than the regular SDSS imaging. This makes \str\ presently the largest deep sky survey in the optical regime.  

\subsection{Stripe82 Stacking}

The SDSS archive provides coadded imaging of \str, but only of the first season of observations (the SDSS \str\ coaddition is explained in \citealp{Annis2011}). We carried out different testing on these coadds. By having obtained surface brightness profiles of our galaxies from these images, it became evident that their coaddition process did not provide imaging with sufficient quality for our purposes. For this reason, we developed our own pipeline to carry out (1) the skysubtraction on the downloaded chips; (2) the quality check of all the available runs; and (3) statistical rejection of low quality runs and the stacking. The pipeline is fully automatic and is mostly written in IDL, but makes use of a few external programs, as well.

All the SDSS observations can be downloaded from the SDSS Data Archive Server (SDSS DAS). Before downloading any image from DAS, we need to decide the size of the stack mosaic that include enough area to extract the surface brightness profiles of our galaxies and estimate any residual sky affecting those. A field-of-view(FOV) of $\sim$10 times the D$_{25}$ proved to be sufficiently large to carry out the profile extraction. To download the images, we extract several information from CasJobs\footnote{http://casjobs.sdss.org/CasJobs/}. SDSS images are identified by their \textit{run}, \textit{rerun}, \textit{camcol}, and \textit{field} numbers. Casjobs returns these numbers corresponding to the position of our target galaxies within a given search radius. (The search radius due to simple geometry is $\sim \sqrt{2} \times$FOV/2) The downloaded \str\ chips are flatfielded and bias-free, but they lack calibration and still contain background. In our pipeline we did not include any calibration process, since what we need is a sufficient number of good quality imaging. We further explain the photometric calibration of the final stacks in Section \ref{sec:photocalib}.

After removing the 1000 counts of the SOFTBIAS added to all SDSS chips, the skyvalue is estimated in the following way. We measure the fluxes in about ten thousand randomly placed, 5 pixel wide apertures. We apply a resistant mean to the distribution of the aperture fluxes, and carry out several iterations to get rid of the fluxes biased by stars, and other background objects. The mean of the bias-free distribution provides accurate measurement of the sky background, and is then substracted from the chips. The skylevel and the background scatter also serve as parameters to take into account when assessing the quality of the imaging run.

After obtaining the mean sky value and scatter of all runs, we study the distribution of these values. To avoid stacking runs with high background, for instance runs observed around full moon, we reject runs where the skylevel is a 3-sigma outlier from the robust mean of all runs. We also need to be careful about not to stack chips with relevant sky gradient, or chips which show signs of detector degradation. To assess skygradients and possible chip anomalies, we use the same scatter information. In case of flat background the skylevel-skysigma relation is expected to conform the same relation for all the runs. For this reason, we fit this skylevel-skysigma relation and reject runs where it lies beyond 1-sigma error of the expected scatter value.

We also would have liked to avoid problems caused by wide-spread PSFs on our final stacks, to this end we only stacked runs with FWHM $\le$ 1.7 arcseconds. To this end, we measured the seeing on every run. We used a  program called PSFEX \citep{psfex} to extract the FWHM. This program makes a robust fit of the PSF by using the stars extracted by SExtractor \citep{Bertin1996}. It is also important to avoid stacking of imaging with high extinction. Since \str\ observations do not come with photometric calibration, we needed to estimate the level of atmospheric extinction in a crude way. We extracted stars from every run, and measured the fluxes in an aperture both on the \str\ run and on the DR7 image. The mean ratio of the measurements computed for all stars in the run was used as a first-order estimate of the extinction: we rejected a particular run, if we found more than 15$\%$ difference of the measured fluxes compared to DR7. Based their histogram, we found that the majority of the runs ($\gtrsim$ 60$\%$) is within this flux difference from DR7, hence this enabled us to create stacks using most of the observations.

Since there have been \str\ observations taken under non-photometric conditions, this kind of quality check is necessary to have a final stack of exceptional quality. Once the non-useful runs were rejected, we start the stacking process which is done in two steps. During the first step, we create shallow mosaics of equal size to the final mosaic from all the fields contained within a given run (or more in case we need to use several camera columns to cover the desired field-of-view). The mosaics are assambled from the skysubtracted chips using SWARP \citep{swarp}. Afterwards, we run SWARP over these shallow mosaics, applying a median stacking to all the available layers. It is important to note that the runs normally have slightly different offsets. Thanks to this, the scattered light from the telescope does not produce any pattern on the images either. Our final stacks do not show any residual sky-gradient or visible sky pattern. Due to the different number of layers used to compose the stacks, the mosaics have slightly different depths. Table \ref{tab:stackprop} shows the number of layers used for the r`-band stacks (in brackets the number of g`-band layers, if that differs by more than 10$\%$). For some of our galaxies, the u'-band and z'-band detector problems prevented us to create reliable mosaics. In these cases (NGC~1068, NGC~1087, UGC~02081, and UGC~02311 in the u'-band and NGC~1068 in the z'-band ) we did not use the u'- and z'-band stacks for further analysis.

\begin{table}[t]
\begin{center}
\caption{\label{tab:stackprop} Stack and profile quality}
\begin{tabular*}{\columnwidth}{@{\extracolsep{\fill}} c|cccc}
\toprule
\textbf{Galaxy} & \textbf{r`-band layers} & T$_{exp}$ (min) & \textbf{ $\mu_{3\sigma}$ } & \textbf{ $\mu_{\pm\sigma-limit}$ } \\
\hline
NGC0450 & (31)42 & $\sim$ 38 & 29.5 & 28.8 \\

NGC0941 & 47 & $\sim$ 42 & 29.9 & 29.4  \\

NGC1068 & 60 & $\sim$ 54 & 30.1 & 29.5 \\

NGC1087 & 55 & $\sim$ 45 & 29.0 & 28.6 \\

NGC7716 & (38)58 & $\sim$ 52 & 29.4 & 28.9 \\

UGC02081 & (43)53 & $\sim$ 48 & 29.8 & 29.2 \\

UGC02311 & 66 & $\sim$ 60 & 29.4 & 28.9 \\
\bottomrule
\end{tabular*}
\end{center}
\end{table}

\subsection{Photometric calibration}
\label{sec:photocalib}

The photometric calibration of our deep mosaics is done by means of comparative photometry. The deep mosaics have an effective exposure time $\sim$ equal to that of SDSS DR7. This makes it straightforward to compare number counts or surface brightness profiles to the DR7 calibrated data. 

The SDSS DR7 photometric calibration is reliable at the level of $\sim$ 2 $\%$. We opted to follow the traditional recipe to determine the zeropoint of our stacks by using the \textit{aa}, \textit{kk}, and \textit{airmass} (the photometric zeropoint, the extinction coefficient and the airmass coefficients). These parameters are provided in the tsField table associated with the chip containing the target object. The surface brightness zeropoint can be given as $-2.5 \times (0.4 \times [aa + kk \times airmass] + 2.5 \times log(t_{exp} \times pixelscale^2)$, where the pixelscale is 0.396''/pixel and the exposure time for each pixel equals to 53.907456 seconds. 

We take the reliable part of the surface brightness profiles of our galaxies obtained from the DR7 imaging and match it to the corresponding part of the \str\ surface brightness profiles. The mean difference is taken as a 'photometric' offset and applied onto the DR7 zeropoint: $ZP_{S82,x} = ZP_{DR7,x} + \Delta \mu_{x}$. Since the the calibration is done in the same photometric filter system, no color term needs to be included.

\subsection{Masking and Profile Extraction}

We extract radial surface brightness profiles on masked mosaics in order to avoid contamination on our light profiles. We apply conservative masking onto sources which clearly do not belong to the galaxy, like foreground stars, background galaxies, etc. These sources are extracted by SExtractor from a master image composed of all 5 bands. This master image is a stack of the different filters scaled to the r-band flux. Sources are extracted in ``cold mode'', optimized for brighter objects. We use some SExtactor parameters such as the measured flux, elongation, and similar to determine the shape and size of these mask regions.  Faint objects (fainter than $\sim 20$ mag) are not masked to statistically correct for the faint foreground stars unresolved from the galaxy image. In some cases, problematic sources, saturated stars, for instance, are masked manually. Since the masking is based on information coming from all 5 filters, we use the resulting mask as a mastermask, and it is applied to all filters. 

In order to extract radial surface brightness profiles corresponding to the underlying outer disk, we need to get representative values of the ellipticity and position angle of this region. Taking a similar approach to SExtractor \citep{Bertin1996}, we do this by computing the second-order moment of the light distribution of the galaxy using the DR7 r`-band image. The second-order moment is directly related to the position angle, the semi-major (A) and semi-minor (B) axis lengths. We fix this ellipticity and position angle for all elliptical apertures. In order to increase the signal-to-noise counteracting its natural decrement due to the lower surface brightness in the outer regime, we used a logarithmic radial sampling with steps of 0.03. In each aperture we estimate the galaxy flux by the 3-$\sigma$ rejected mean of the pixel values of that aperture. This helps to minimize the effect of morphological features like a spiral arm crossing the aperture. 

\begin{figure*}[t]
\centering
\includegraphics[width=0.9\textwidth]{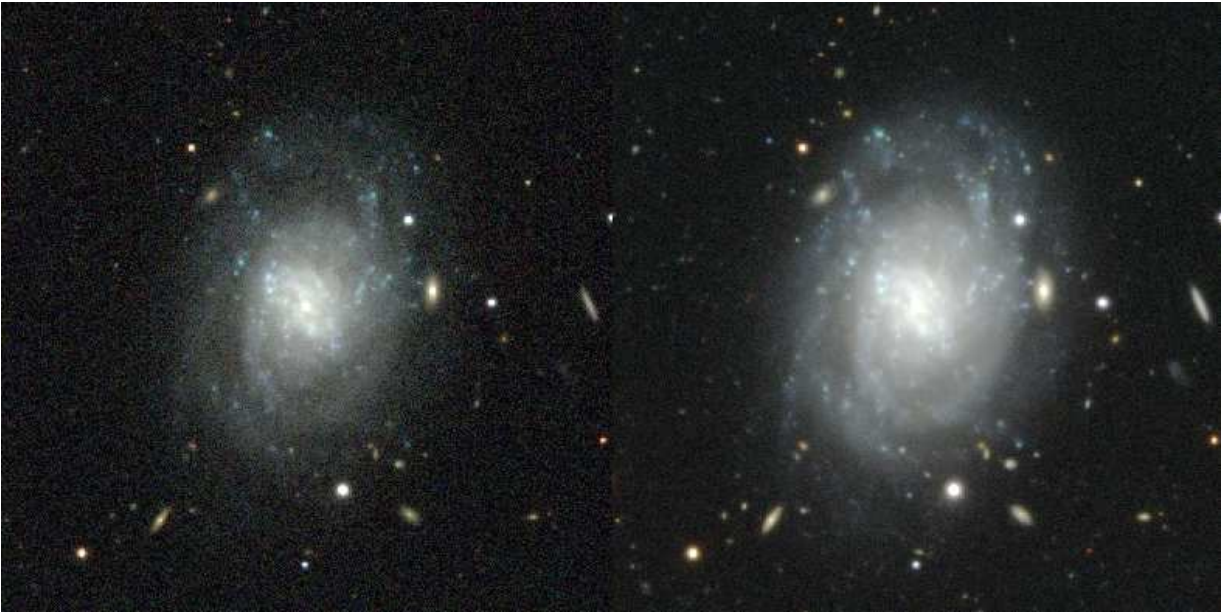}
\caption{\textit{Left}: RGB image of NGC~0941 created using the standard (single exposure) SDSS DR7 g`-,r`-,and i`-band imaging. \textit{Right}: RGB image of NGC~0941 created using the g`-,r`-,and i`-band stacks of \str\ exposures of NGC~0941. The \str\ imaging is $\sim$ 2 magnitudes deeper than the standard SDSS imaging, this effectively means that S/N $\sim$ 1 on the SDSS DR7 imaging corresponds to features of a surface brightness equal to 25 mag arcsec$^{-2}$, however the same signal-to-noise on \str\ imaging corresponds to 27 mag arcsec$^{-2}$.}
\label{fig:dr7vss82_1}
\end{figure*}

Our light profile, however, can still be contaminated by some residual sky background. This background is estimated by using equally spaced apertures. We obtain the number count profile of the galaxy up to very far distances, and we chose a large aperture where the profile becomes flat, beyond the identifiable profile of the galaxy. The 3-$\sigma$ rejected mean of the fluxes inside this aperture gives a robust estimate of the residual sky background (for further information, see \citetalias{PT06}, their Section 3.3). The error of the background determination is key to decide the surface brightness level down to which we trust our profiles. Following a conservative approach, this is placed where the profiles obtained by either over- or undersubtracting the sky measurement by $\pm\sigma$ start to deviate with more than 0.2 mag from the original profile. To see the limiting surface brightness of our profiles in the r'-band see Table \ref{tab:stackprop}.

\subsection{Data quality}

After stacking, \str\ imaging is expected to be $\sim$ 2 magnitudes deeper than the standard SDSS imaging. This amount of difference can be observed directly on the imaging: spiral galaxies show more outer structure, for instance, revealing outer disks extending farther out, meanwhile the inner morphology remains unchanged. As an example, see the DR7 and \str\ RGB color composits of NGC~0941 on Figure \ref {fig:dr7vss82_1}.

A signal-to-noise $\sim$ 1 per pixel on the DR7 imaging corresponds to features with a surface brightness of $\sim$ 25 mag arcsec$^{-2}$. However, by summing up thousands of pixels in the outer apertures of the surface brightness profiles, we can reach S/N $\sim$ 5 at the level of $\sim$ 27 mag arcsec$^{-2}$. This is the S/N required to assure errors $\leq$ 0.2 mag. The same S/N $\sim$ 1 per pixel on the \str\ imaging can be associated with the surface brightness level of $\sim$ 27 mag arcsec$^{-2}$.

\citetalias{PT06} showed that with standard SDSS imaging it is possible to extract radial surface brightness profiles down to $\mu_{r`} \sim$ 27 mag arcsec$^{-2}$, hence we expect to be able to obtain reliable profiles down to $\mu_{r`} \sim$ 29 mag arcsec$^{-2}$. Table \ref{tab:stackprop} shows that this depth is fairly stable for all our galaxies, however, in case of NGC~1068 or NGC~0941, due to a larger number of pixels falling in the outer apertures or smaller intrinsic noise thanks to better sky conditions, we were able to extract its profile down to $\mu_{r`} \sim$ 30 mag arcsec$^{-2}$. 

These deep surface brightness profiles also prove the high quality of the regular SDSS imaging. We made a direct comparison between the r`-band profiles of NGC~0941 in Figure \ref{fig:dr7vss82_2}.

\section{Analysis}

\subsection{Surface brightness profiles}

We obtained surface brightness profiles in all the five SDSS filters. These profiles are $\sim$ 2 magnitudes deeper than regular SDSS profiles, giving way to explore the stellar populations of disks down to $\sim$ 0.5 M$_{\sun}$ pc$^{-2}$. The deepest profiles, however, are obtained in the g'- and r'-filters, this is due to the better detector/filter response of these filters which results in smaller instrinsic noise. Due to this we explored in detail the r'-band surface brightness profiles and the $(g'-r')$ color. Our results can be found in the Galaxy Atlas for every galaxy separately (see Appendix \ref{sec:galaxy_atlas}). 

\subsubsection{Classification}
\label{sec:class}
We used our deep r'-band surface brightness profiles to classify the surface brightness profiles of our galaxies. We followed the scheme presented in \citetalias{PT06} and \citet{Erwin2008}. The surface brightness profiles of Type~I galaxies are described by one single exponential, meanwhile Type~II and Type~III profiles both exhibit a well defined feature called the break. After an exponential decline, Type~II profiles are followed by a downbending-, Type~III profiles with an upbending-exponential after this inflection point. There exist several sub-classes of Type~II profiles, related to the possible origin of the break. We distinguish the so called \textit{Outer Lindblad Resonance} breaks (Type~II-ORL), for these galaxies the break in the surface brightness profile lies close to the resonance zone caused by the bar, at $\sim$ 2 times the bar radius. In case the break cannot be associated with such a well-defined phenomena, it is classified as \textit{Classical Truncation} (Type~II-CT). We show the results of our classification in Table \ref{tab:diskprop}.

\begin{figure*}[t]
\centering
\includegraphics[scale=0.7,angle=90.]{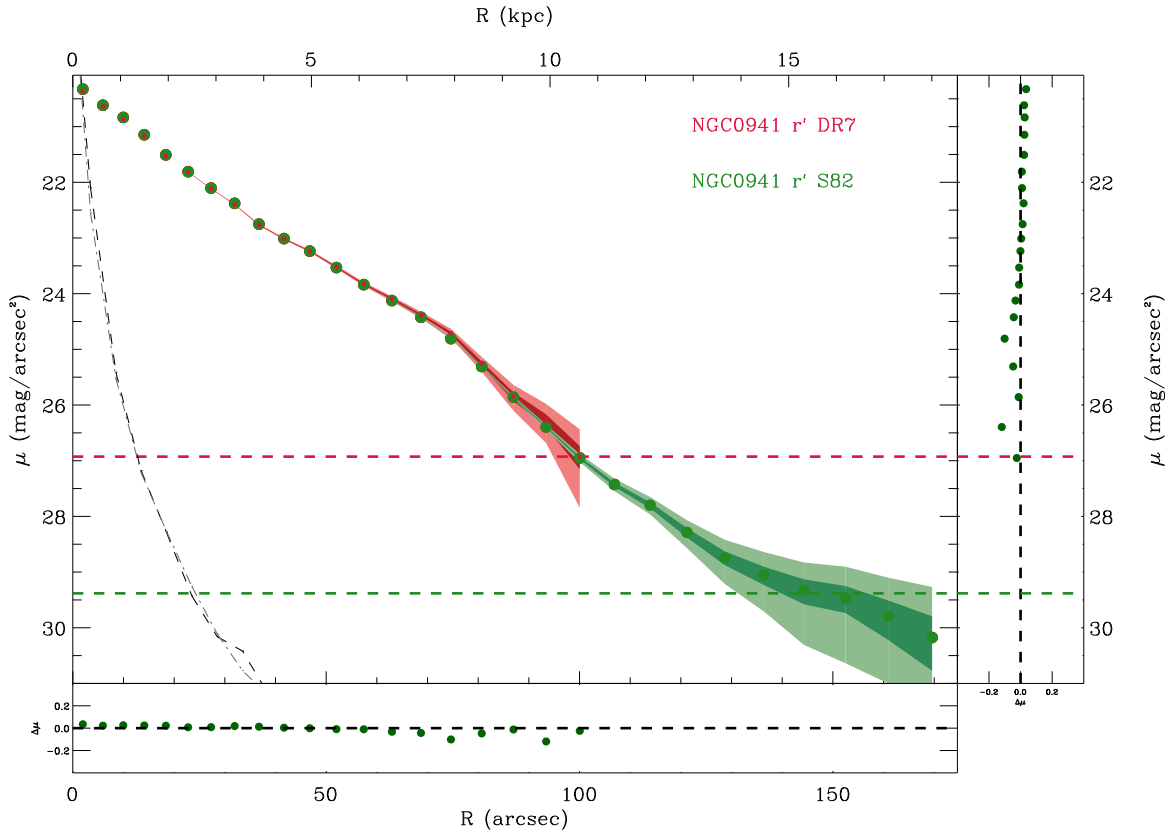}
\caption{Comparison of the surface brightness profiles of NGC~0941 obtained from standard SDSS DR7 imaging and \str\ stacks. In red we plot the surface brightness profile obtained from the standard SDSS imaging, the \str\ profile is overplotted in green. The shaded regions indicate the 1 and 3$\sigma$ under- and over-substracted profiles in both cases. The surface brightness limits of our profiles are placed where the 1$\sigma$ under- and over-substracted profiles start to deviate by more than 0.2 magnitude from the original profile. The DR7 limiting surface brightness (26.9 mag arcsec$^{-2}$) is indicated by the red dashed line, the \str\ limiting surface brightness (29.4 mag arcsec$^{-2}$) is indicated by the green dashed line. On the joint panels we plotted the deviation of the DR7 profiles from the \str\ profile against the surface brightness and radius, this deviation is well below 0.2 magnitudes. This is a further proof of the reliability of the standard SDSS imaging. \textit{Stripe82 PSF}: we also draw the PSF profiles of the g'- and r'-band imaging scaled to the r'-band central flux of NGC~0941. We observed no anomalous color gradient of the PSF.}
\label{fig:dr7vss82_2}
\end{figure*}

\subsubsection{Excess of light at $\mu_{r'} \sim$ 28 mag arcsec$^{-2}$}

The surface brightness profiles of our seven galaxies can be fitted into the aforementioned basic scheme. However, we note that most of our galaxies exhibit some extra light above the outer disk expectations starting at $\sim$ 28 mag arcsec$^{-2}$. This extra light follows a flattened profile that we think is not associated with any morphological feature belonging to the outer disk. In fact, in this paper, we work under the hypothesis that this extra light is diffuse light from the stellar halo. This extra light appears and shows the same structure in the other deep SDSS bands (g' and i'), as well.

To further check our hypothesis, we investigated the origin of these features. First, we had to be sure that no artifact due to observational errors could possibly affect the quality of our profiles. There are several known problems related to integrated photometry like flat-fielding, skysubtraction, contribution by saturated stars or extended PSFs which could cause artificial excess (or defect) on our light profiles.

\paragraph{Flatfielding} SDSS imaging due to their drift-scan technique used during the observations show no evidence of flatfielding issues. 

\paragraph{Skysubtraction} On our surface brightness profiles we already took into account the possible effect of an error in the estimation of the backgound. We drew shaded regions corresponding to the 1$\sigma$ and 3$\sigma$ under- and oversubtracted profiles (see Appendix \ref{sec:galaxy_atlas}). The turnup point of the stellar halo feature (typically at $\mu_{r'} \sim$ 28 mag arcsec$^{-2}$) is at a surface brightness level $\geq$ 15$\sigma$ above the skylevel. The fact that we also see the same behavior for all the galaxies and in different bands, reinforces the idea that this is a real feature.

\paragraph{Saturated stars} There is only one galaxy, NGC7716, where there is significant contribution from a saturated star that makes the interpretation of our surface brightness profile harder. This, however, could be minimized by extensive masking over the galaxy disk itself. We assessed the size of the region affected by scattered light by obtaining a pseudo surface brightness profile of this saturated star. We used similar approach to check the effect of saturated stars in case of other galaxies. 

\paragraph{Scattered light from the bulge} Given the depth of our images, we might expect the effect of the PSF on the bulge (the sharpest feature) of our galaxies to have a significant contribution at low surface brightness levels in form of an extended wing. This could cause the bulge light to scatter far enough to affect the surface brightness profiles in the outer regions. To check the shape of the PSF at low surface brightness levels, we need to model accurately the PSF of the stacks. We used again PSFex to create the models of the PSF of our stacks. These PSF models are usually $\sim$ 12 magnitude deep and might extend over $\leq$ 50 pixels. By scaling these PSF models to the central surface brightness of the galaxy (see Figure \ref{fig:dr7vss82_2}), we have seen that the extention and depth of the PSFs is negligible and could not affect the outer disk profiles. (We also considered the possibility of the bulge emerging above the disk, as a cause to observe the upturns on these profiles. Some arguments have been made in recent years that Type~III profiles, that exhibit an excess of light over the disk in the outer regions, can be associated with the bulge emerging above the disk. The features we observe on the surface brightness profiles are a bit akin to Type~III behavior, although in late-type spirals it would be somewhat unexpected to find bulges reaching $\sim$ 10 kpc or farther out.) 

After checking the above issues, we are confident that the excess of light observed at $\mu_{r'} \sim$ 28 mag arcsec$^{-2}$ is real. We compared the surface brightness levels of our galaxies to several nearby galaxies like Andromeda (M31), M33, M81, or NGC2403. These galaxies are known to have well studied faint structures which were identified by several independent groups using 'resolved star' technique \citep[e.g.][]{Ferguson2002,Ferguson2007,Ibata2007,McConnachie2009,Barker2009,Barker2012}. The base of our comparison is that our galaxies have quite similar stellar masses. \citet{Courteau2011} showed that the regime of stellar halos actually starts at the level of $\mu_I \sim$ 27 mag arcsec$^{-2}$ (using the color predictions\footnote{See also on http://miles.iac.es} of \citealt{Vazdekis2010} for old and metal-poor stellar populations, this magnitude translates into $\mu_{r'} \sim $27.5 \sbdim) corroborating our assumption of observing the signs of stellar halos on our profiles.

\subsubsection{Deriving disk and stellar halo parameters}
\label{sec:decomp}

In addition to the classification of the surface brightness profiles, we carried out a detailed structural analysis on the observed r'-band profile. To derive reliable parameters of the disk and stellar halo, a Bulge/Disk/Stellar Halo decomposition was done by simoultaneously fitting each component. With this we also avoid overestimating the light contribution of the different components, especially that of the stellar halo to the total galaxy light. We fitted a composite model built up from functions corresponding to the known or assumed shapes of the different morphological components observed on the profile. It has been known for some time that disks are well fitted with exponential profiles \citep{Patterson1940,Freeman1970}. However, the majority of disks does not follow a single exponential profile, but rather exhibits light deficiency or in some cases excess in the outer disks \citep[\citetalias{PT06}]{vanderKruit79,Erwin2005,Erwin2008} after a well identifiable point of the profile, which is often called the \textit{break radius}. Due to these reasons, we modeled the disk by a double-exponential to take into account the existence of these breaks:
\begin{equation}
I(R) = I_{0,1} e^{R/h_1} \Theta(R_{br}) + I_{0,2} e^{R/h_2} (1 - \Theta(R_{br})).
\end{equation}

The central intensity of the inner and outer disks are I$_{0,1}$ and I$_{0,2}$, and $\Theta(R_{br})$ is the step function:
\begin{equation} 
\Theta(R_{br}) = 
\begin{cases}
1 & \text{if } R \leq R_{br} \\
0 & \text{if } R > R_{br}. 
\end{cases}
\end{equation}

The step function assures the double-exponential profile with \textit{h$_1$} and \textit{h$_2$} scalelengths before and after \textit{R$_{br}$} break radius.

The bulges of spiral disks are modeled with an r$^{1/n}$ profile \citep*{Andredakis1995}. The 1D S\'ersic r$^{1/n}$-profiles \citep{Sersic1968} can be written as follows:

\begin{equation}
I(R) = I_{e}exp \left\{  -b_n  \Bigg[ \Bigg(\frac{R}{R_{e}}\Bigg)^{1/n} -1 \Bigg]  \right\},
\end{equation}

where $R_e$ is the effective radius, and $I_{e} = I(R_e)$. The quantity b$_n$ is a function of the shape parameter n, and is chosen so that the effective radius encloses half of the total luminosity, a good approximation can be given as $b_n \sim 2n - 0.324$ \citep[e.g.][]{Graham2001}.

Stellar halo profiles obtained by star-counts method are often fitted by S\'ersic-profiles, as well, \citep[see e.g.][]{Courteau2011}. 

Altogether, the composite profile has 10 free parameters that needs to be fitted. In addition to this, in the flat, outer region where the stellar halo dominates the profiles, the observational errors are playing a significant role. Due to this, and after visual inspection, we decided to minimize the uncertainty in determining the properties of the stellar halo by fixing its shape with a S\'ersic index n = 0.5, that is a Gaussian profile. \citet{Trujillo2002} showed that the mass density profile of the $r^{1/n}$ family can be extremely well approximated by an analytical expression (see their Eq. 7), which yields an exact case for n = 0.5. By choosing this constraint therefore we ensure a moderately declining mass distribution, on the other hand allowing a higher value for the S\'ersic-index would mean that the stellar halo has a prominent ``over the disk component'' in the inner parts of the galaxy (see their Fig. 2). 

The profile fitting is performed with the \textit{MPFIT} routine in the Interactive Data language (IDL), this routine is based on the Levenberg-Marquardt algorithm \citep{Markwardt2009} to carry out non-linear least-squares minimization. Although this fitting method is extremely sensitive to initial conditions, with a proper set of input parameters the algorithm gives excellent results.

The decomposition of the light profile yields robust estimation of the scales of different galaxy component, like disk scalelengths and halo effective radius, that can be directly compared to theoretical expectations, and it also becomes possible to compare these halos to that of the Milky Way or M31. In case of the Local Group galaxies it becomes possible to integrate the observed \textit{starcount} profile out to large radii, obtaining a robust and model-free extimation on the contribution of the halo light to total galaxy light. In our case, we only trace the inner part of the stellar halo, hence we cannot estimate the halo contribution by simple integration. However, one advantage of using S\'ersic models (as it is shown e.g. in \citealt*{Trujillo2001}) is that the global properties, such as the total light (L$_{T}$) and the mean effective surface brightness ($\langle \mu \rangle_e$), can be given in the form of exact analytical formula based on the aforementioned three observables ($n$, $R_e$, and $I_{e}$): 

\begin{equation}
L_{T} = \textit{k}_L I_{e} R_{e}^{2}, 
\end{equation}

where \textit{k}$_L$ is the ``structural parameter'' that depends on the S\'ersic index: 
\begin{equation}
\textit{k}_L = e^{b_n} \frac{2\pi n}{b_n^{2n}} \Gamma(2n), 
\end{equation}

and

\begin{equation}
\langle \mu \rangle_e \equiv -2.5 log \frac{L_{T}}{2\pi R_{e}^{2}}. 
\end{equation}

\begin{figure*}[t]
\centering
\includegraphics[width=0.9\textwidth]{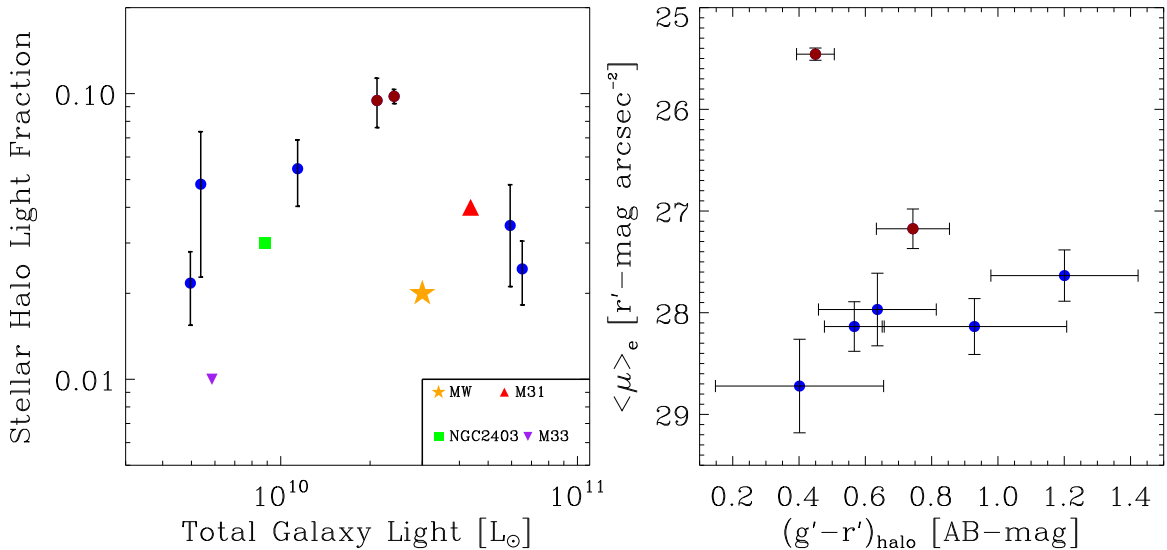}
\caption{\textit{Left panel}: Fraction of the stellar halo light vs. the total galaxy light. The stellar halo light is estimated by fitting with a n=0.5 S\'ersic-mode the region of the stellar halo of the galaxies, meanwhile to obtain the total galaxy light we integrated the observed profile. The dark red points coincide with the galaxies where we identify the presence of tidal features (\textit{NGC7716}) or minor mergers (\textit{NGC1087}). The stellar halo light fraction observed in the Milky Way \citep{Carollo2010}, M31 \citep{Courteau2011}, M33 \citep{McConnachie2010} and NGC2403 \citep{Barker2012} are also overplotted. The light contribution of the stellar halo to the total galaxy light varies from $\sim$ 1$\%$ to $\sim$ 5$\%$, but in case of ongoing mergers the stellar halo light fraction can be as high as $\sim$ 10$\%$. \textit{Right panel}: Effective surface brightness of the stellar halo vs. the integrated (g'-r') color of the stellar halo. When computing the stellar halo color, the halo region is taken outside the radius where the halo light start to contribute with $\geq$ 50$\%$. We find colors ranging $(g'-r') \sim $ 0.4 -- 1.2. The extremely red halo colors ($g'-r' > $0.8) are found in galaxies with no evident interaction.}
\label{fig:haloprop}
\end{figure*}

The evaluation of the total galaxy light is as follows: we integrate the observed profile until the last measured point. Outside this region (to take into account the stellar halo contribution) we extrapolate the behavior of the fit of stellar halo to infinity. The stellar halo light fractions of the 7 galaxies compared to the host galaxy luminosity are shown in Figure \ref{fig:haloprop}.

\subsection{Stellar Population Analysis} 
\label{sec:analysis_stpop}

\subsubsection{Color Profiles}
\label{sec:analysis_color}

In addition to the light distribution of these late-type spirals we obtained color profiles to explore the radial distribution of stellar populations in our galaxies. Normally, color profiles cannot be studied down to the same surface brightness level as the surface brightness profiles, due to the fact that errors propagate quadratically and the uncertainty of the color is necessarily attributed to the uncertainty of the sky estimation in both filters. The errors coming from the sky estimation, however, can be alleviated by using large bins. We find that the most convenient strategy is to place the color bins around the well identified morphological features such as the bulge, disk etc. Nevertheless, it is important to avoid using a single bin to describe regions where the color gradient is steep, and it would mean mixing very different stellar populations. The bins we applied for the g'-r' color profiles (see Appendix \ref{sec:galaxy_atlas}) are a compromise between the increase of signal-to-noise and underlying color gradient. 

In the regime of the stellar halo we observe no significant color gradient, therefore to give a robust result, we obtained a single (g'-r') color for our stellar halos using one large bin. To derive these colors for all stellar halos in a consistent way, we placed the inner boundary of the color bin at the radius (R$_{>50}$) where the stellar halo light starts to dominate the total light, and the outer boundary at the radius where the surface brightness profile reaches the 3$\sigma$ skynoise. This way the binsize depends on the stellar halo properties and instrinsic noise properties of the image. 

We also obtained color profiles for other adjacent filters (u`-g`,r`-i`,i`-z`), in total providing 4 color profiles that can be studied in detail. 

The Galactic extinction has been taken into using the Schlegel et al. (1998) values. We used the IDL code provided by Schlegel et al.\footnote{http://astro.berkeley.edu/$\sim$marc/dust/index.html} to get the values corresponding to the different galactic positions of our sample. 

\subsubsection{Color errors}

Deriving colors for the faint stellar halo calls for a lot of care. At very faint surface brightness levels, the colors are affected by the shape of the PSF of the different bands. Anomalous PSF scattering can cause excess of light in one band, hence causing an artificial color. This effect was presented e.g. by \citet{deJong2008} who accurately proved that (1) PSF tails on stacked SDSS imaging can extend as far as $\sim$ 100 pixels reaching $\sim$ 14 magnitudes below the central surface brightness, (2) color gradients and reddening could be artificially caused by the extended scattering especially in the SDSS i'-band. 

Particularly, in the case of edge-on galaxies \citep{Zibetti2004,Zibetti2004a}, according to the \citet{deJong2008} results, the scattered light from the central disk contaminates the light of the stellar halo. In face-on projections, however, the artificial color gradient induced by the different wing extentions of the PSF plays a less relevant role. This is for two reasons: (1) the slope of the surface brightness profiles in face-on projections are much moderate than in the edge-on case being less affected by the PSF; (2) whereas the stellar halos are found after the disk inmediately in edge-on projections, in face-on projections the transitions are much moderate and the PSF size is not large enough to affect the stellar halo in their outer regions.

Colors could also be affected by an inaccurate sky subtraction. To account for the propagation of the uncertainty of the sky estimation into the color, we over- and undersubtracted the galaxy images with 1$\sigma$ uncertainty, and obtained total intensity, hence magnitudes in the different bins from those images. The mean magnitude difference $\langle\delta\rangle$ = $|$m$_{+\sigma}$ - m$_{-\sigma}|/2$ was propagated quadratically: $$\Delta = \sqrt{\langle\delta\rangle_{g}^2 + \langle\delta\rangle_{r}^2}$$ and used as error on the color.


\subsubsection{Stellar surface mass density profiles}
\label{sec:ssmd}
To study the stellar mass distribution of our spirals, we computed the stellar mass density (\ssmd) profiles using the \citet{Bell2003} prescriptions for the mass-to-light (M/L) ratio.

In this work we assume a Kroupa IMF \citep{Kroupa2001}, which according to \citet{Bell2003} implies a deduction of 0.15 dex from the mass-to-light ratio using the following expression:\\
\begin{equation}
 log_{10} (M/L)_{\lambda} = \left(a_{\lambda}\ +\ b_{\lambda}\ \times\ color \right)- 0.15, 
\end{equation}
where for the SDSS (g'-r') color $a_{\lambda} = -0.306$ and $b_{\lambda} = 1.097$
is applied to determine the r' band $M/L$.

Once we know the $(M/L)$ ratio, using the expression below it is straightforward to link the stellar mass density (\ssmd) profile with the surface brightness profile at a given wavelength ($\mu_{\lambda}$).
\begin{equation}
 log_{10} \Sigma_{\star} = log_{10} (M/L)_{\lambda} - 0.4(\mu_{\lambda} - m_{abs,\odot,\lambda}) + 8.629,
 \label{sigma}
\end{equation}
where $m_{abs,\odot,\lambda}$ is the absolute magnitude of the Sun at wavelength $\lambda$, and \ssmd\ is measured in ${\rm M}_{\sun}{\rm pc}^{-2}$.

\subsection{Stellar population models}
\label{sec:vazdekis}

Comparing the observed stellar halo color to the predictions of single stellar population (SSP) models we expect to get a crude hint on the stellar population that might build up the observed stellar halos. Since we lack a wide multiwavelength data, we opted for constraining the observed (g'-r') colors. We used the latest MILES SSP models. These empirical models are based on real stellar spectra and give excellent coverage both for metallicity and ages \citep{Vazdekis2010}. Another important aspect of using these models is that they provide predictions for different IMFs.

We have decided to constrain the stellar populations of the stellar halos in a novel way. This was necessary since the age-metallicity degeneracy prevents us to gain a deeper understanding of these stellar populations, moreover we would like to see how the observational errors affect the measurements of age and metallicity. To this end we created synthetic color maps based on the Vazdekis models. 

\begin{figure*}[t]
\centering
\includegraphics[angle=90,width=0.85\textwidth]{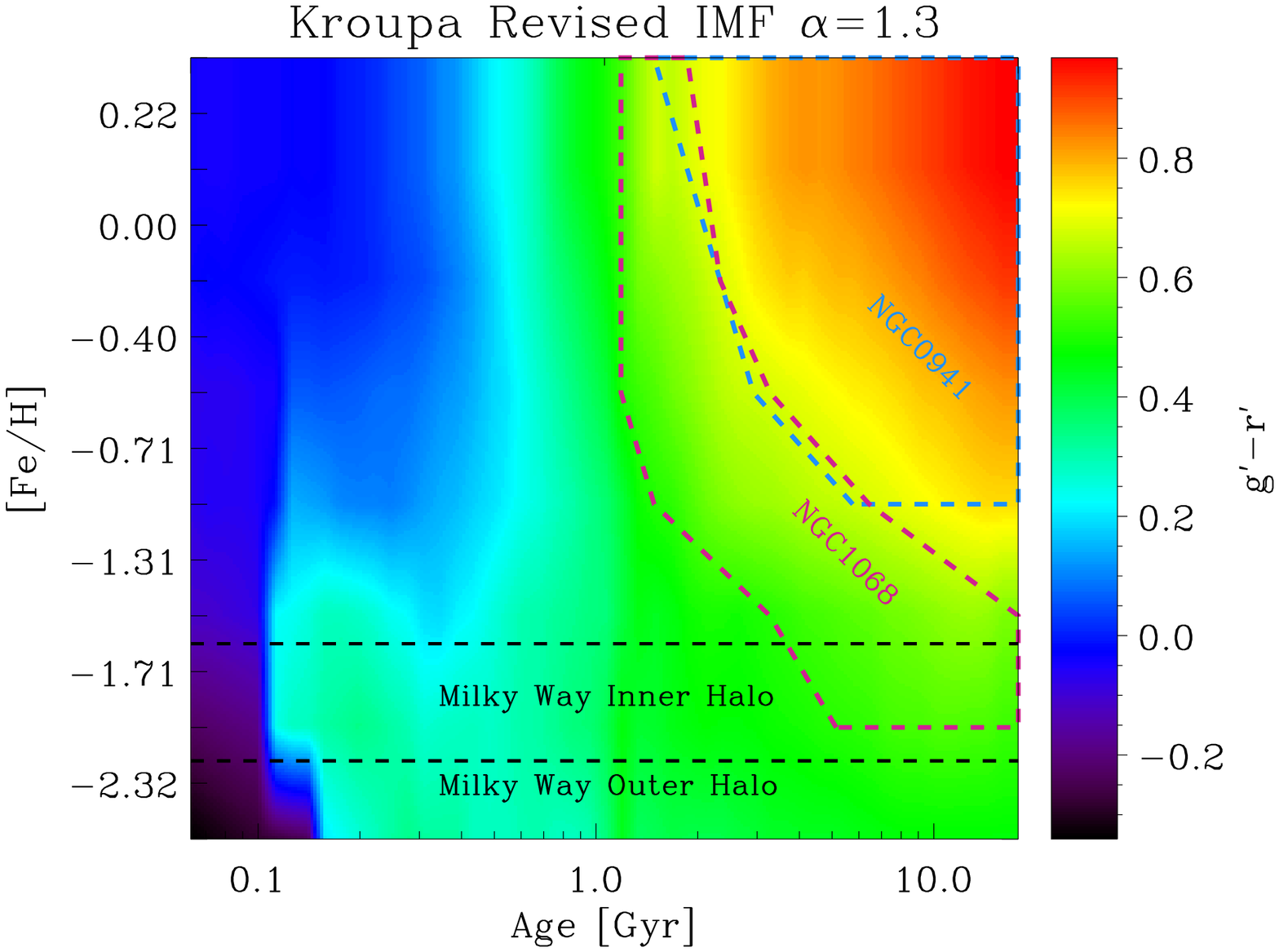}
\includegraphics[angle=90,width=0.85\textwidth]{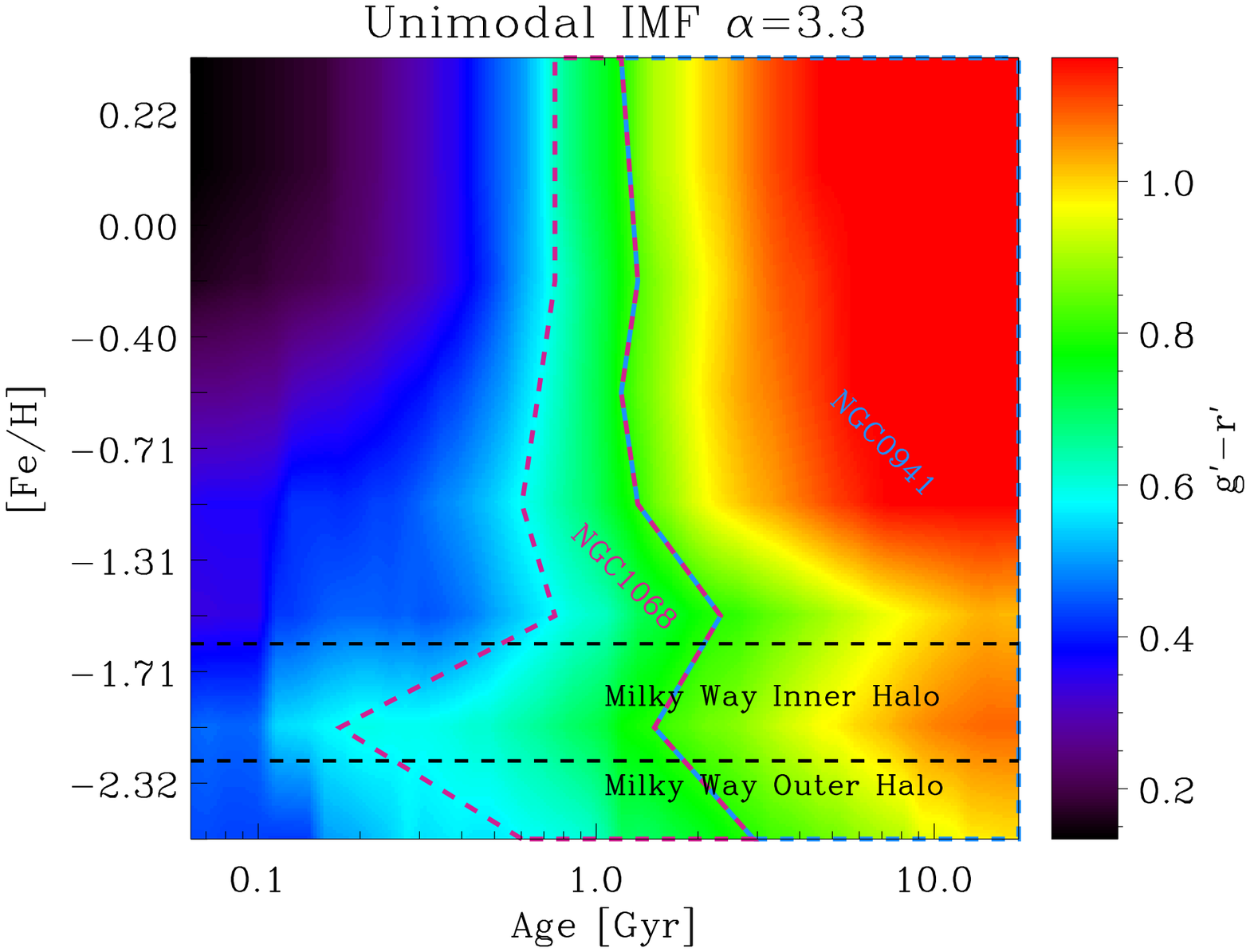}
\caption{\textbf{Metallicity-Age maps}: Using the MILES SSP models \citep{Vazdekis2010} to get a hint on what stellar populations might build up the stellar halos. We illustrate the expected age and metallicity for two different galaxies. In case of NGC~0941 the stellar halo is extremely red and models produced by regular Kroupa IMF are unable to reproduce this color. The blue (NGC~0941) and magenta (NGC~1068) dashed lines show the age and metallicity solutions that are compatible with the observed colors within their error bars. The upper panel displays the model colors for Kroupa IMF and the bottom panel with a bottom-heavy IMF.}
\label{fig:imf}
\end{figure*}

As a first step, we took all SSP model spectra and convolved them with the SDSS g' ad r' filters in order to compute model colors for each age and metallicity. Then we composed a two-dimensional map of the model colors and placed our measurements over the age-metallicity grid. To visualize how the age-metallicity degeneracy appears in the colors of the stellar population, we assigned the colors of the rainbow table in IDL to the values computed from the models. The rainbow colors change smoothly from violet(black) to red. Instead of computing $\chi^2$ values from the observed colors, we decided to use the observed colors and its errors, directly on these maps. The range of colors within the observed color and its error bar can be associated with a range of ages and metallicities, creating by default a two-dimensional problem. This is shown in Figure \ref{fig:imf}.

We have found that some of our observed stellar halo colors are so red, that they lay outside the grid of models based on the conventional Kroupa IMF with slope $\alpha$ = 1.3. The observed halo colors are redder than the reddest colors of those models. Redder model colors can be expected of models based on bottom-heavy IMFs \citep{Vazdekis2010}. Although, there is a wide range of IMF slopes between $\alpha=[0.3,3.3]$ available, here, we simply took the model that provided us with the reddest colors, therefore we used models with Unimodal IMF with a slope of $\alpha = 3.3$. To comprehend how the IMF affects the intrinsic color distribution, see Figure \ref{fig:imf}.

\section{Results}
\subsection{Surface brightness profiles}

The surface brightess profiles of our 7 galaxies have been classified accordingly to the scheme explained in Section \ref{sec:class}. 

Our sample contains one Type~I disk NGC~1068(=M77). In this work, NGC~1068 has been classified as Type~I (in \citetalias{PT06} it was classified as Type~II-ORL). We considered here that rather than having a break associated with the Outer Lindblad Resonance cited by \citetalias{PT06}, the extended oval distortion in the disk accounts for the extra light over the disk observed on the surface brightness profile (see Figure \ref{fig:NGC1068_apx}). NGC~0450, NGC~0941, and UGC~02081 exhibit Type~II classical truncations, the break of UGC~02311 lies by the resonance zone caused by a bar, hence it is classified as Type~II-ORL. When classifying NGC~1087 and NGC~7716, we had to be a bit cautious. These galaxies seem to have a break feature in the inner disk that can be associated with a centrally more concentrated star forming region. This however is not the kind of phenomena we designate as Type~II, hence these galaxies are classified as Type~II*. In their outer disk both galaxies exhibit Type~III up-bending. In NGC~7716 this extended Type~III feature has a disky origin starting at about 80 arcseconds at a level of 25 mag arcsec$^{-2}$, meanwhile in NGC~1087 the up-bending shows a spheroidal origin. To be consistent with previous classification scheme, these 2 galaxies have been classified as Type~II+III. 

Interesting to note that the high contribution of the stellar halo light in case of NGC~1087 explains why this galaxy in \citetalias{PT06} was classified as Type~III. However, in this sample, we find that the presence of the stellar halo causes a ``Type~III like'' feature on all profiles. 

Beyond the classification, we derived several properties related to the radial light distribution of these galaxies. Applying the decomposition explained in Section \ref{sec:decomp}, we inferred several structural properties for the bulge, disk, break, and stellar halo. 

\subsubsection{Bulge properties}

This paper was not aimed to discuss bulges in detail, however, it is interesting to note that the bulges of our galaxies are well fitted with low S\'ersic-index($n$) values $n \lesssim$ 1.5 (see Table \ref{tab:bulgeprop}). \citet*{Kormendy2004} discussed that small $n$ values are signatures of pseudo-bulges. There is also a general trend noted by \citet*{Andredakis1995} that in late-type (Sc and later) galaxies the bulge profiles are very close to being pure exponentials. We observe the same trend for our objects, and that indeed in some cases, like NGC~0941, we have alternatively confirmed the presence of a pseudo-bulge (Kormendy, private communication).

\begin{table}[t]
\begin{center}
\caption{\label{tab:bulgeprop} Bulge Properties}
\begin{tabular*}{\columnwidth}{@{\extracolsep{\fill}} c|ccc }
\toprule
\multicolumn{1}{c}{\textbf{Galaxy}}
& \textbf{n}
& \textbf{R$_{e,b}$}
& \textbf{$\mu_{e,b}$} \\
\multicolumn{1}{c}{} 
& \multicolumn{1}{c}{}
& \multicolumn{1}{c}{\textbf{[kpc]}}
& \multicolumn{1}{c}{\textbf{[mag arcsec$^{-2}$]}} \\
\hline

NGC0450 &   0.95 &   0.94 & 20.77 \\
NGC0941  & 0.64  & 1.34 & 21.73 \\
NGC1068  & 0.92  & 0.54 & 16.54 \\
NGC1087  & 0.73  & 0.24 & 17.81 \\
NGC7716  & 1.50  & 0.60 & 18.18 \\
UGC02081  & 0.50  & 0.45 & 20.49 \\
UGC02311  & 1.33  & 4.18 & 20.21 \\

\bottomrule
\end{tabular*}
\end{center}
\end{table}


\begin{table*}[ht]
\begin{center}
\caption{\label{tab:diskprop} Disk and Break Properties}
\begin{tabular*}{\textwidth}{@{\extracolsep{\fill}} cc|ccccccccccc }
\toprule
\multicolumn{1}{c}{\textbf{Galaxy}}
& \multicolumn{1}{c}{\textbf{SB-Type}}
& \multicolumn{3}{c}{\textbf{R$_{break}$ }}
& \multicolumn{2}{c}{\textbf{h$_{in}$}}
& \textbf{h$_{in}$}
& \textbf{$\mu_{0,in}$} & \textbf{$\mu_{0,out}$} & \textbf{$\mu_{br}$} 
& $(g'-r')_{br}$ & $\Sigma_{br}$ \\
\cmidrule(r){3-5}
\cmidrule(r){6-7}
\cmidrule(r){9-11}
\multicolumn{1}{c}{} 
& \multicolumn{1}{c}{} 
& \multicolumn{1}{c}{\textbf{[``]}} & \textbf{[kpc]} & \textbf{[\textit{h}$_{in}$]}
& \multicolumn{1}{c}{\textbf{[``]}} & \textbf{[kpc]}
& \textbf{[\textit{h}$_{out}$]}
& \multicolumn{3}{c}{\textbf{[mag arcsec$^{-2}$]}} 
&  & [$M_{\odot} pc^{-2}$] \\
\hline
 NGC0450 & Type~II-CT & $ 82.3$ & $ 9.7$ & $ 2.20$ & $ 37.4$ & $ 4.4$ & $ 2.21$ & $21.19$ & $18.29$ & $23.53$ & $0.26$ & $ 8.2$ \\
\vspace{0.05cm}
 NGC0941 & Type~II-CT & $ 71.5$ & $ 7.6$ & $ 3.45$ & $ 20.7$ & $ 2.2$ & $ 1.72$ & $20.82$ & $18.13$ & $24.50$ & $0.35$ & $ 4.2$ \\
\vspace{0.05cm}
 NGC1068 & Type~I &  \nodata  & \nodata & \nodata  & $ 71.5$ & $ 5.3$ & \nodata & $20.91$ & \nodata & \nodata & \nodata & \nodata \\
\vspace{0.05cm}
 NGC1087 & Type~II*+III & $ 51.4$ & $ 5.2$ & $ 1.37$ & $ 37.4$ & $ 3.8$ & $ 1.65$ & $19.77$ & $18.81$ & $21.17$ & $0.39$ & $98.8$ \\
\vspace{0.05cm}
 NGC7716 & Type~II*+III & $ 39.2$ & $ 6.9$ & $ 2.53$ & $ 15.5$ & $ 2.7$ & $ 1.10$ & $19.85$ & $19.59$ & $22.47$ & $0.40$ & $30.8$ \\
\vspace{0.05cm}
UGC02081 & Type~II-CT & $ 66.5$ & $11.8$ & $ 3.36$ & $ 19.8$ & $ 3.5$ & $ 2.22$ & $21.60$ & $17.16$ & $25.32$ & $0.28$ & $ 1.7$ \\
\vspace{0.05cm}
UGC02311 & Type~II-ORL & $ 42.1$ & $21.0$ & $ 2.38$ & $ 17.7$ & $ 8.8$ & $ 2.30$ & $21.47$ & $18.11$ & $23.92$ & $0.43$ & $ 8.8$ \\
\hline
\bottomrule
\end{tabular*}
\end{center}
\end{table*}

\subsubsection{Disk and Break properties}

Our model fitting provides robust estimations of disk properties, such as inner, and outer scale-lengths, break radius, inner and outer disk central surface brightness, and the break surface brightness. The results of the fitting can be found in Table \ref{tab:diskprop}. A detailed statistical analysis of these parameters is beyond the scope of this paper, due to the low number of observed galaxies in our sample. However, on a one-by-one basis it is interesting to compare their properties to previous studies of late-type disks, and later on, to put this in context with the observed stellar halo properties.

\citetalias{PT06} studied $\sim$ 90 galaxies where they have found that $\sim$60$\%$ of late-type disks contain Type~II breaks. Based upon this result it is not surprising that four of our seven galaxies ($\sim$60$\%$) exhibit Type~II behavior. \citetalias{PT06} also studied the typical radius where surface brightness breaks usually occur, taking into account that the morphologically distinct sub-classes might have very different origins. Five of our galaxies (NGC~0450, NGC~0941, NGC~1068, NGC~1087 and UGC~02081) were part of the PT06 sample, as well. Besides of the re-classification of NGC~1068 and NGC~1087, the break radii we measured for NGC~0941(R$_{br} \sim$ 72$''$) and UGC~02081(R$_{br} \sim$ 67$''$) are different from the PT06 values (R$_{br}=$ 82$''$ and 53$''$, respectively). Part of these differences are due to the different ellipticities we applied when obtaining the surface brightness profiles. Different ellipticity means different scale-length properties, as well. However, differences in the measurement of the scale-length might also arise due to a more accurate Bulge/Disk decomposition method we used to infer disk properties. We nevertheless are in position to say that statistically the obtained break properties are within the PT06 expected ranges. In light of both results, we conclude that the method applied in PT06 was sufficient to give contraints on the disk structure on a statistically significant level, and that the use of shallow DR7 data does not compromise disk measurements.

\begin{figure}[Ht]
\centering
\includegraphics[scale=0.45]{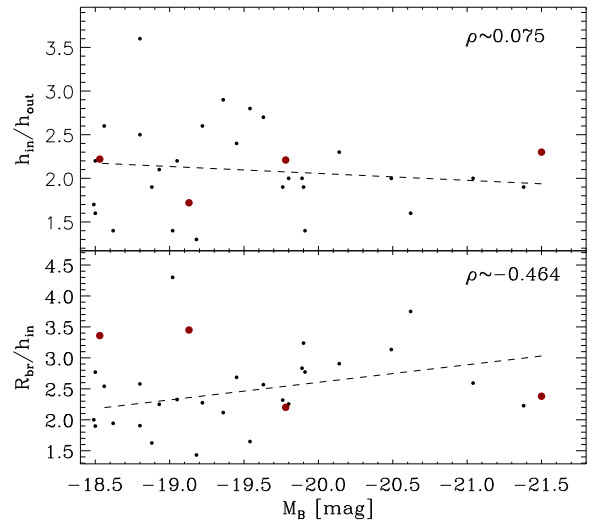}
\caption{Disk property correlations with absolute magnitude for the \citetalias{PT06} Classical Truncation sample. The red dots are the properties obtained for the Type~II galaxies of the current sample. Overplotted in both panel are robust linear fits to guide the eye and in the upper right corner the Spearman rank correlation coefficient. \textit{Upper panel}: The inner and outer disk scale-length ratio for the \citetalias{PT06} sample of Type~II-CT galaxies. \textit{Bottom panel}: The break radius in inner scale-length units.}
\label{fig:diskprop}
\end{figure}

Similarly to \citetalias{PT06}, we characterized the strength of the break by the ratio of the inner and outer scale-lengths. The mean ratio of the inner and outer scale-length on the PT06 sample was 2.5$\pm$0.5. They also reported a lower range of values that could be as low as 1.1 (case of NGC7716). To compare how our measurements fit into the \citetalias{PT06} relations, we overplotted the $\frac{h_{in}}{h_{out}}$ of this paper with the PT06 values (see the upper panel of Figure \ref{fig:diskprop}). 

\citetalias{PT06} showed that Type~II Classical Truncations (NGC~0450, NGC~0941 and UGC~02081) are expected to occur at $\sim9\pm2.4$ kpc, meanwhile breaks related to resonances (UGC~02311), show a wider span of galactocentric distances with the same mean distance ($\sim$9 kpc). Both distributions showed the tendency of larger break radius occuring in higher luminosity galaxies (see their Figure 7). Taking into account their results, it is not surprising to measure such a large break radius in UGC~02311, since this galaxy is $\sim$2-3 magnitudes brighter than the other galaxies with Type~II-CT breaks (see Table \ref{tab:sample}). The break radius measured in relation to the scale-lengths, however, does not reveal such differences between the two Type~II sub-types, its typical value being 2-3 \textit{h$_{in}$}, this maches well with the mean parameter value of 2.5$\pm$0.6 found in \citetalias{PT06} (see lower panel of Figure \ref{fig:diskprop}). The inner scale-lengths of these galaxies correspond to the typical values measured in local galaxies \citep{deJong1996,MacArthur2003}, although UGC~02311 has a scale-length larger than usual ($\sim$ 9 kpc).   

The inner disk central surface brightnesses are  $\sim$ 0.5 magnitude fainter compared to the PT06 values. This difference could be attributed to the better treatment of the light contribution from the bulge done in this paper. The wider spread deviations in the outer disk central surface brightnesses compared to the PT06 values are more difficult to explain. The proper fitting of the stellar halo should in any case cause the outer disk scale-length to shorthen and its central surface brightness to brighten, resulting in a sharper break. However, \citetalias{PT06} breaks appear to be sharper (having brighter central surface brightnesses, and larger $h_{in}/h_{out}$), probably the effect of the ellipticity, that we intend to investigate in our upcoming paper (BT2012, in prep.). 

In order to investigate how Type~II break properties (namely the inner and outer scale-length ratio) depend on the observing wavelength, we carried out a Bulge/Disk decomposition on the profiles in all 5 filters (if available) by applying the bulge and disk part of the fitting routine explained in Section \ref{sec:decomp}. To be fully consistent, we fitted the same disk region for all filters for our galaxies. We have cut the profiles taking care of not mixing the regions where the stellar halo light would have a significant contribution. During the fit we fixed the break radius. 

\begin{figure}[b]
\centering
\includegraphics[angle=90,width=0.45\textwidth]{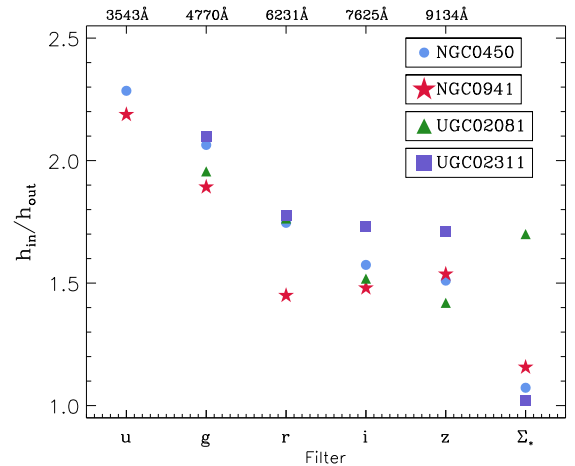}
\caption{The strength of the break (indicated with the inner-outer disk scale-length ratio) obtained from the surface brightness profiles observed in the 5 SDSS bands and from the stellar surface mass density profile (See \ref{sec:ssmd}). The inner-outer scale-length ratio decreases rapidly towards the redder filters, moreover on the stellar surface mass density profile the break disappears (h$_{in}$/h$_{out} \sim $1). However, in case of UGC02081 the break remains prominent on the stellar surface mass density profile, implying that the break in this galaxy is entirely different phenomena and reseambles more of the cut-offs found by \citet{vanderKruit79}.}
\label{fig:break}
\end{figure}

We find that the $h_{in}/h_{out}$ decreases smootly from u'- to z'-bands in our Type~II galaxies (see Figure \ref{fig:break}).

\subsubsection{Stellar Halos}

\begin{table*}[t]
\begin{center}
\caption{\label{tab:haloprop} Properties of stellar halos}
\begin{tabular*}{\textwidth}{@{\extracolsep{\fill}} c|ccccccc }
\toprule
\multicolumn{1}{c}{\textbf{Galaxy}}
& \multicolumn{1}{c}{\textbf{R$_{e}$ }}
& \multicolumn{1}{c}{\textbf{$\langle\mu\rangle_{e}$ }}
& \multicolumn{3}{c}{\textbf{R$_{turnoff}$}}
& \multicolumn{2}{c}{\textbf{R$_{>50}$}}\\
\cmidrule(r){4-6}
\cmidrule(r){7-8}
\multicolumn{1}{c}{}
& \multicolumn{1}{c}{\textbf{[kpc]}}
& \multicolumn{1}{c}{\textbf{[mag arcsec$^{-2}$]}} 
& \multicolumn{1}{c}{\textbf{[``]}} & \textbf{[kpc]} & \textbf{[\textit{h}$_{in}$]}
& \multicolumn{1}{c}{\textbf{[``]}} & \textbf{[kpc]}\\
\hline
 NGC0450 &  18.90 &  27.64 & 128.68 &  15.22 &   3.44 &  152.48 &  18.04 \\
 NGC0941 &   9.90 &  28.14 & 100.05 &  10.62 &   4.83 &  121.20 &  12.87 \\
 NGC1068 &  38.00 &  28.14 & 393.82 &  29.21 &   5.51 &  513.78 &  38.11 \\
 NGC1087 &  13.50 &  25.46 & 111.46 &  11.19 &   2.98 &  159.11 &  15.97 \\
 NGC7716 &  35.00 &  27.74 &  86.96 &  15.39 &   5.61 &  106.89 &  18.91 \\
UGC02081 &  20.11 &  28.72 &  84.08 &  14.88 &   4.25 &  97.36 &  17.23 \\
UGC02311 &  40.00 &  27.97 &  62.98 &  31.39 &   3.56 &  74.62 &  37.19 \\
\bottomrule
\end{tabular*}
\end{center}
\end{table*}

Using SDSS \str\ data gives us the opportunity to study the stellar content of disk galaxies down to a very deep level $\sim$30 mag arcsec$^{-2}$ in the r'-band. This depth enables us to get a glimpse of the stellar halo on the surface brightness profiles: the light contribution from the stellar halo leaves a well identifiable upturn on the r'-band profiles, and when the stellar halo starts to dominate the light the profile becomes flatter. We observe this flattening on the profiles of all 7 galaxies that helped us to establish that the typical surface brightness level at which stellar halos start to affect the shape of the surface brightness profiles is $\sim$27 mag arcsec$^{-2}$. 

Although, this is a purely observational concept, it is interesting to discuss the radius where the light profile of the (outer) disk of the galaxy \textit{first} starts to deviate from the exponential decline. We refer to this radius as the turn-off radius (R$_{turnoff}$) in Table \ref{tab:haloprop}. The estimation of this radius (and surface brightness) is model-dependent. Based on the assumption that the disk continues exponentially without having an actual edge, a 0.5 magnitude change (larger than the error bars of our measurements at these levels) could be expected if the halo light contribution reaches $\sim$ 25$\%$ that of the disk. This 0.5 mag deviation from the expected exponential shape of the surface brightness profile is well observable. The measured values of R$_{turnoff}$ fall at 10 to 30 kpcs from the center of the galaxy, in terms of the inner scale-length between 3 and 5.5 $h_{in}$. 

These upturns should not be confused with the outer regions where the stellar halo clearly dominates the light of the galaxy (since around the upturn the disk is still the major contributor of the galaxy light). The radius where more than 50$\%$ of the galaxy light comes from the stellar halo is referred as R$_{>50}$. At this radius the stellar halo contribution clearly overcomes that of the disk and dominates the flat appearance of the surface brightness profile. As it is seen in Table \ref{tab:haloprop}, there are, on average, 2-3 kpcs differences between R$_{turnoff}$ and R$_{>50}$.

In all these aspects, NGC~1087 is a special case. In this galaxy the stellar halo starts to affect the shape of the surface brightness profile at higher surface brightness levels. Our results imply that NGC~1087 has an unusually bright stellar halo with $\langle \mu \rangle_e \sim$ 25.5 mag arcsec$^{-2}$. Probably this is the reason why this galaxy, based on shallow surface brightness profiles (see \citetalias{PT06}), was classified as Type~III. This altogether raises an important question: just how many (late-type) galaxies classified as Type~III are in reality Type~I galaxies with a bright stellar halo?

The mean effective surface brightness ($\langle \mu \rangle_e$) of our stellar halos have a typical value of $\sim$ 28 \sbdim\ (see Table \ref{tab:haloprop}). NGC~1087 and NGC~7716, the two interacting galaxies have brighter \sbeff\ (see Table \ref{tab:haloprop}). 

\begin{figure*}[t]
\centering
\includegraphics[width=0.9\textwidth]{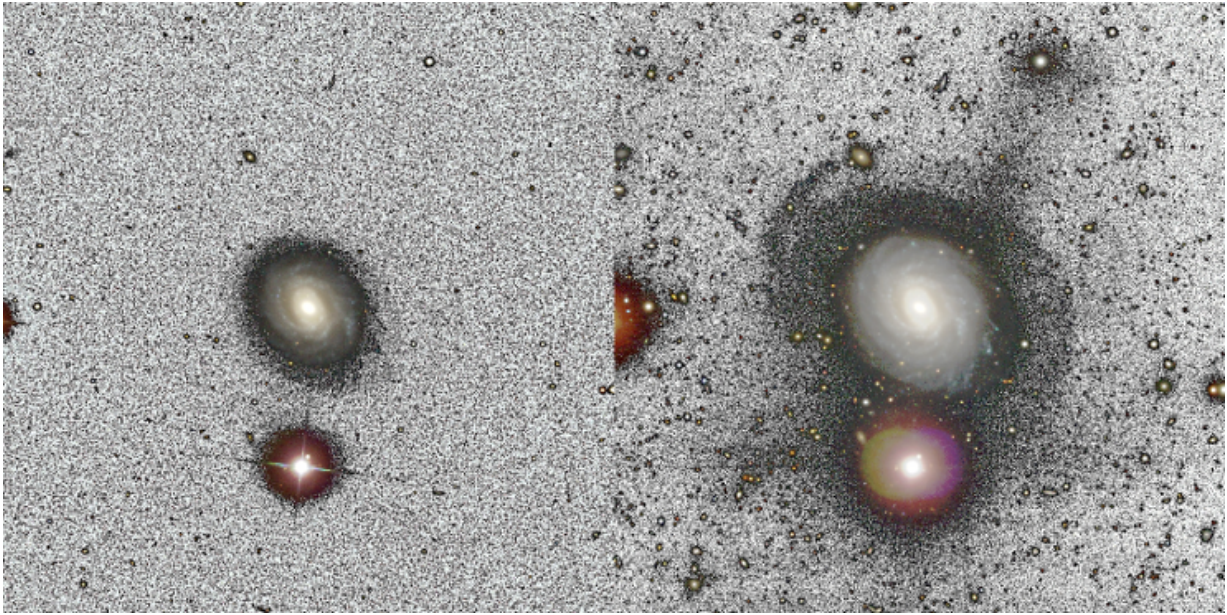}
\caption{\textit{Left image}: NGC7716 in SDSS r'-band (single exposure). The limiting surface brightness of this image is $\sim$27 mag arcsec$^{-2}$, which corresponds to 3$\sigma$ of the skynoise. The skynoise (similarly to \citetalias{PT06}) is the resistant mean of the skyvalues measured on the pixels in an aperture surrounding the galaxy. \textit{Right image}: \str\ coaddition of NGC7716 with clearly visible tidal streams likely originating from two satellites. The surface brightness of the streams is $\sim$27 mag arcsec$^{-2}$ in r'-band. }
\label{fig:tidal}
\end{figure*}

An important aspect of our study is the light fraction contained within the stellar halo compared to the total stellar light of the galaxy. We find fractions in the range of 1--5$\%$. These are typical values observed in nearby galaxies, as well. In two galaxies (NGC~1087 and NGC~7716), though, the light fraction of the stellar halo reaches nearly $\sim$ 10$\%$. It is interesting to note that in both galaxies we note signs of ongoing interaction or merging. NGC~7716 is a clear-cut example of tidal streams surrounding a galaxy (see Figure \ref{fig:tidal}), that appears as a sharp feature on the surface brightness profile, and has a surface brightness of $\sim$ 26.5 mag arcsec$^{-2}$.

There could be several candidates as progenitors for this tidal stream(s), but due to lack of spectroscopic data we cannot decide whether the objects along the stream do indeed belong to the system or are only background objects. There is only one known satellite of this galaxy reported \citep{Zaritsky1993}, but that is found too far (at a projected distance of $\sim$ 130 kpc) to be connected to the streams. NGC~1087 is not such an obvious case of interaction, but a misaligned outer disk might indicate that the equilibrium of the disk has been disturbed recently by an external source.

\subsection{Color profiles}

The u',g',r',i',and z' filters of SDSS give a full coverage of the Spectral Energy Distribution (SED) in the optical regime. In principle, using all the five filters gives us the opportunity to learn whether the surface brightness distribution and color gradients exhibit any kind of dependence on the observing wavelength allowing us to constrain the stellar populations responsible for the energy output in the different galactic components. Unfortunately, not all bands behave as well as g' and r'. Stacking u'-, i'-, and z'-band data did not result in such exceptional images and we cannot derive more than some exploratory conclusions. Due to this, we mainly focused on g'- and r'-band data.

We can observe several interesting behavior on the published surface brightness and color profiles. Although, on this sample it is not possible to derive statistically robust correlations, we find that the behavior of the surface brightness and color is connected to the surface brightness types of these galaxies. 

In the Type~II-CT galaxies in this sample (like NGC~0450 and NGC~0941) we find that the Type~II behavior appears more pronounced in the u'-band. In NGC~0450 the u'-band surface brightness profile seems to have a sharp cut-off (see Figure \ref{fig:NGC0450_apx}), causing an enhanced ``U-shape'' on the (u'-g') color profile. Towards redder wavelengths the break feature seem to weaken and the ``U-shape'' flattens out (see Figures \ref{fig:NGC0450_apx},\ref{fig:NGC0941_apx})

In \citetalias{BTP08} we obtained robust color profiles of Type~I, Type~II and Type~III galaxies. For individual galaxies the (g'-r')color profiles reached a depth coinciding with $\sim$ 25 mag arcsec$^{-2}$ on the r'-band surface brightness profiles. In the particular case of Type~II galaxies, when we combined the information of many objects, we were able to derive a mean color profile down to $\sim$ 27 mag arcsec$^{-2}$. This mean color profile was not just deeper but predicted the \textit{general} behavior of Type~II galaxies. Due to that analysis we found that this type of galaxies exhibit a ``U-shape'' on their (g'-r') color.

Now, using \str\ data we got the opportunity to trace the color profiles down to fainter levels and confirm for instance the classical ``U-shape'' in case of individual galaxies. NGC~0450 and NGC~0941 are typical examples of this Type~II behavior. Their g'-r' color profiles (see Figures \ref{fig:NGC0450_apx}, \ref{fig:NGC0941_apx}) follow nicely the ``U-shape'', although, the gradient in the inner disk in NGC~0941 is much milder but with a prominent reddening beyond the break. The same striking reddening can be observed beyond the break in NGC~7716 and UGC~02311. UGC~02081, despite of being clasified as Type~II based on its surface brightness profile, exhibits no reddening beyond the break. Actually, its flat color profile resambles that of what we found to be intrinsic in Type~I galaxies. The color profile of NGC~1068 (Type~I) is quite flat, it has some minor reddening around the turn-off radius presumably caused by the stellar halo light contribution. The (r'-i') color profile of NGC1068 does not reach the same depth, still, around the turn-off radius it shows the same slight reddening.  

The color profile of NGC~1087 (see Figure \ref{fig:NGC1087_apx}) is puzzling. Based on its surface brightness profile we classified this galaxy as Type~II*, associating the apparent break in the inner disk to the presence of a more concentrated star forming region. Meanwhile, the g'-r' color profile exhibits a fairly deep ``U-shape'' around that break followed with a flattening around the radius where the stellar halo contribution starts to dominate the light. The overall shape of this color profile in \citetalias{BTP08} was associated with Type~III galaxies.

Due to the low number of galaxies ($\sim$ 10) we did not obtain robust color profiles for Type~I galaxies in \citetalias{BTP08}, although, we have found the same flattened behavior for the mean color profile as along the disk in NGC~1068, the only Type~I disk we could study in detail in this work. NGC~1068 has a rather dusty disk, so it is not unexpected that the mean color of the disk of NGC~1068 is rather red (g'-r' $\sim$ 0.6). 

In \citetalias{BTP08} we found a strong correlation for Type~II galaxies between the color at the break with the absolute magnitude of the galaxy. If we compared the color at the break (see in Table \ref{tab:diskprop}) to the the observed (g'-r') = 0.47 $\pm$ 0.02 mag mean color of \citetalias{BTP08}, we should expect that the main reason why the majority of the observed colors at the break are bluer than the \citetalias{BTP08} mean color is that these systems bear with less total stellar mass. It is indeed true that lower-mass systems such as NGC~0941 and UGC~02081 exhibit the bluest colors at the break in this sample. UGC~02311 also exhibits a blue color at the break, despite its high stellar mass. Although, in case of this Type~II-ORL break one could expect enhanced star formation around the resonance zone. The presence of star forming rings around the outer Lindblad Resonance has been observed in many spirals \citep[e.g.][]{Buta1991,Buta2002,Grouchy2010}.  The two peculiar Type~II galaxies, NGC~1087 and NGC~7716, are similar in stellar mass to the mean value of \citetalias{BTP08}, but only NGC~7716 has (g'-r') color close to the BTP08 mean value.  

\subsection{Stellar surface mass density}

As explained in Section \ref{sec:analysis_color}, we obtained $M/L$ ratio profiles from (g'-r') colors. These profiles were converted into stellar surface mass density (\ssmd) profiles that serve as a proxy of the stellar mass distribution of our 7 galaxies. Though, when reaching the stellar halo the Bell \& de Jong conversion formula \citep{Bell2001} seems to break down, and the density basically becomes independent of the radial distance. This phenomena is quite evident in case of NGC~0450 and NGC~0941 (see e.g. Figures \ref{fig:NGC0450_apx}, \ref{fig:NGC0941_apx}). It is important to note that in these two galaxies we have observed extremely red colors in their stellar halo. If these red colors are due to some extreme stellar populations (with a different IMF, for instance), the \cite*{Bell2001} empirical formula cannot give valid predictions for the \mtol, since it is largely based on the assumption of a Universal Salpeter-like IMF.

The \ssmd\ profiles of Type~II galaxies do not follow the shape of the r'-band surface brightness profiles. Similarly to the findings of \citetalias{BTP08}, the break diminishes on the stellar surface mass density profile. To quantify the 'sharpness' of the break on the \ssmd\ profile, we applied the same Bulge/Disk decomposition on the \ssmd\ profiles as for the surface brightness profiles. Not surprisingly, the inner-outer scale-lengths ratio takes a value of $\sim$ 1 in NGC~0450, NGC~0941 and UGC~02311. In UGC~02081 the break, however, remains quite prominent, implying a different phenomena in this galaxy. (See last column on Figure \ref{fig:break}).

In \citetalias{BTP08} we found that at Type~II galaxies, the mean stellar mass density at the break radius is $\sim$ 10 \ssmddim. Taking into account that at the break positions of NGC~1087 and NGC~7716 the stellar surface mass density is much higher than this value ($\sim$99 \ssmddim\ and $\sim$31 \ssmddim, respectively), not classifying these galaxies as Type~II-CT seems justified. On the \ssmd\ profile of UGC~02081 we observe quite an irregular cut-off, that seems to be a real mass drop after the break radius.  Also, the \ssmd\ has a low value at the break radius (\ssmd$_{,br}$ = 1.7 \ssmddim), this value is much lower than the average value found in \citetalias{BTP08}.

\subsection{Stellar populations in the stellar halos}

The observed (g'-r') colors for our stellar halos cover a large range from $\sim$ 0.4 to $\sim$ 1.2. This is a rather unexpected variety of colors and stellar populations contained within the stellar halos, independently of the total stellar mass of the host galaxy. 

In order to learn about the nature of the stellar populations behind these observed colors, we compared them to stellar population model predictions \citep{Vazdekis2010}. As it is already explained in Section \ref{sec:vazdekis}, Figure \ref{fig:imf} is only a crude way to visualize what the stellar population content of these stellar halos might be. We chose two typical stellar halos to do this analysis, the red stellar halo of NGC~1068 ($(g'-r') \sim 0.55$) and the extremely red stellar halo of NGC~0941 ($(g'-r') \sim 0.95$). It is important to note, that the observed stellar halo colors are redder than that of the disk or even the bulge (with the exception of NGC~1068, which has a very dusty disk, hence the stellar halo appears to be bluer than the disk).

It is relatively obvious that with the error bars of our measurements neither metallicity, nor age can really be constrained. However, by assuming that the stellar halo properties of our galaxies are similar to the ones in our neighborhood, we can make some further but tentative constraints based on observations of nearby galaxies. For instance, it is widely accepted that the Milky Way stellar halo is dominated by a metal-poor population ($[Fe/H] \lesssim -1.5$) \citep[e.g.][and references therein]{Ryan1991,Carollo2007,Ivezic2008}. In other nearby galaxies, like M31 \citep{Ferguson2002,Chapman2006,Ibata2007}, M33 \citep{Davidge2003,McConnachie2006,Barker2007,Grossi2011}, or NGC~2403 \citep{Davidge2003,Barker2012}, the stellar halos consist of predominantly old and metal-poor ($[Fe/H] \sim$~-0.7~--~-1.5) stellar populations. 

It is interesting to note that SSP models produced with conventional Kroupa IMF ($\alpha = 1.3$) predict that the stellar halo color observed in NGC~0941 would imply a stellar halo that is fairly metal-rich and old. This is at odds with the measurements in nearby galaxies. By changing the IMF, though, it is possible to look for solutions that would attribute the red colors to metal-poor stellar populations. It has already been discussed in the literature \citep[e.g.][and references therein]{Zackrisson2006} that a bottom-heavy IMF could be responsible for such red colors. Following this idea, we applied a bottom-heavy IMF with a slope of $\alpha = 3.3$ to compute our model (g'-r') colors. These models indeed yield possible solutions of $\tau \gtrsim$ 4 Gyr old and $[Fe/H] \lesssim$ -1.5. In case of NGC~1068, this IMF would imply a young stellar halo (younger than in case of Kroupa IMF) with any metallicity possible. 

\section{Discussion}

Using deep SDSS \str\ data we observed the stellar content of 7 late-type galaxies down to $\mu_{r'} \sim$ 30 mag arcsec$^{-2}$. This is roughly a stellar surface mass density of $\sim$ 0.1 M$_{\sun}$ pc$^{-2}$, allowing us to study the stellar halo component of these galaxies.

At present, Stellar halos are best studied by resolved star technique, and there have been only a few attempts to study stellar halos by integrated photometry \citep[e.g.][]{Jablonka2010}. However, to explore galaxies beyond our neighborhood, integrated photometry is needed. Resolved star technique provides ample information on stellar halos: the surface brightness of stellar halos in external galaxies, and how their contribution affects the surface brightness profile. From recent observations of nearby galaxies such as M31 or NGC2403 we know the structure of the disks and stellar halos of these galaxies in unprecedented detail \citep[e.e.][]{Courteau2011,Barker2012}. 

Tha galaxies in our sample are simillar in stellar mass and in Hubble-type to many of these well studied systems. Although, stochastic variations in the evolutionary path of these individual galaxies could cause striking differences as to how the structural properties of the different substructures vary from galaxy to galaxy on the same mass scale. Still, \citet{Courteau2011} argued that one should expect to encounter the signatures of stellar halos in surveys that reach below the depth of $\mu_{r'} \sim $27.5 \sbdim. Using SDSS data \citetalias{PT06} extracted surface brightness profiles down to a similar depth, but the questioned remained: \textit{What is the real extent of disks? Does the exponential disk continue down to lower surface densities as predicted by some models?} In many cases the structure of the outer disk does not change right below 27 mag arcsec$^{-2}$. The outer disk indeed follows an exponential surface brightness profile down to the level where there is a smooth continuation into the region where the stellar halo light dominates over the disk light (causing the upturns discussed beforehand at the level of $\mu_{r'} \sim$ 28 \sbdim). 

Due to this mixture of components, we are unable to determine the real extent of the disks. It would seem that this disk-stellar halo light conspiracy is due to the observing geometry. In face-on geometry we cannot decouple the light of the stellar halo from that of the disk. Because when we study the surface brightness distribution of these two components, we rely on the superimposed two dimensional projection of both the stellar halo and the disk. The only way to explore further the disk light would be in edge-on projection. In edge-on projection by integrating the light of the galaxy along the line-of-sight of the disk, the contrast between the disk and the stellar halo is enhanced, allowing us to explore around $\sim$2 mag deeper the disk component.

Our deep surface brightness profiles has modified our view on how surface brightness profiles should be classified. By making use of, for instance, high quality color profiles it was possible to rethink the previous classification scheme, especially in relation to Type~III profiles.

\subsection{On the nature of Type~III profiles}

The nature of Type~III profiles since their discovery \citep{Erwin2005} proved to be contradictory. Any comprehensive picture regarding the nature of Type~III profiles needs to explain: why we observe two sub-class of Type~III profiles (with disk like or spheroid like isophotes in the outskirts), the origin of the peculiar color profile of Type~III galaxies, and why the statistics of Type~III profiles decreases towards later Hubble-types. 

The fraction of Type~III profiles becomes larger in early-type spirals \citep[\citetalias{PT06},][]{Erwin2008}. Consequently, as the prominence of the bulge increases with the same trend, one could hazard that Type~III profiles are nothing else but the bulge component emerging above the disk component on the surface brightness profiles. This would also explain the roundish isophotes that many of these Type~III galaxies present in their outer regions. Indeed, being motivated by this idea, \citet{Maltby2012} studied the connection between the bulges and Type~III profiles in a sample of intermediate redshift galaxies. They found that especially in early-type disk galaxies Type~III profiles are caused by the aforementioned bulge emergence, but they could not refute the idea that in many cases, however, it remains a pure disk phenomenon. \textit{What is the origin of Type~III profiles in late-type galaxies where the bulge component is rather small to be held responsible for such phenomena, and where the light of the outer region clearly follows disky isphotes?}

In our sample, we have found that the contribution of the stellar halo light on the surface brightness profiles causes a similar upturn as on 'classical' Type~III profiles. In case of NGC~1087 this contribution is so high that it starts to affect the surface brightness profile at a higher surface brightness level. Due to this, this galaxy previously was classified as Type~III \citepalias{PT06,BTP08}. It is natural to ask whether it is a unique behavior of this galaxy or a general rule that would mean that we should find many examples once looked carefully. Interestingly, M81 (a twin of NGC~1087 with its absolute magnitude of $M_B$ = -20.7 mag) is also a system with an unusually bright halo component, though \citet{Barker2009} also consider the possibility that the extended bright component in M81 belongs to a disk feature (e.g. thick disk). Regardless of this, the light fraction of this component compared to the total luminosity in both galaxies is $\sim$ 10 $\%$. In NGC~7716 the stellar halo is similarly bright, moreover in this galaxy the misaligned outer disk, that causes a sharp Type~III break, can be associated with clear interaction. The tidal stream we identified around this galaxy is clearly connected to this misaligned outer disk. 

NGC~1087 and NGC~7716 have a central star forming region in their inner disk that causes a high surface brightness break on their surface brightness profile that we indicated with an ``asterix Type~II'' profile (distinguishing from the usual Type~II classification because this break is \textit{not} the phenomena we usually associate with the \textit{end} of the spiral arms). We can assume that the enhanced star formation in the inner disk is triggered by the recent interaction that has either supplied the material necessary to form stars or redistributed the gas already settled inside the disk. If this scenario is plausible, we can inmediately understand why we observe such a peculiar shape of the Type~III color profiles. As it was already shown in \citetalias{BTP08} (see their Figure 1) Type~III galaxies seem to retain the ``U-shape'' at shorter galactocentric radii, with a red ``plateau'' residing at the Type~III break followed by a blueing towards the outskirts. It is remarkable that NGC~1087 retains this shape on its (g'-r') color profile, at the Type~II* break with the bluest color and at the Type~III break with a red ``plateau''. NGC~7716 also exhibits the bluest color at the position of the Type~II* break.

Combining all the information from the surface brightness and color profiles, we propose the following unified picture of Type~III (late-type) galaxies: Type~III profiles are not an individual class of galaxies but rather, Type~I or Type~II galaxies with a stellar halo. The surface brightness level at which the stellar halo contributes depends on the recent evolutionary history (and stellar mass) of the galaxy. Galaxies are transformed by (recent) interaction. The infalling new material enriches the stellar halo, can add extra material to the disk and trigger star formation leading to a Type~II+III mixed profile. Meanwhile, a more quiescent history means a less developed stellar halo, that would explain the existence of pure Type~III profiles (see the published profiles of \citetalias{PT06} and \citealt{Erwin2008}). 

Due to this, in any flux limited survey that cannot probe the surface brightness levels of quescient stellar halos, the ``Type~III profiles'' would remain unobserved. This also would explain the low number of Type~II+III profiles in the \citetalias{PT06} sample ($\sim 5\%$), that necessarily comes from the combined effect of a high surface brightness Type~II break and a bright stellar halo/tidal stream on the surface brightness profile. Type~II+III late-type galaxies would be late disks 'caught' during a merger event.

\subsection{Further clues on outer-disk formation}

\textit{What clues can we deduce from the results of this paper in relation to the formation of breaks in the surface brightness profile of disk galaxies? Can these deep surface brightness, color and stellar surface mass density profiles constrain the formation of these individual systems by making use of the predictions of current models?}

The \textit{formation of Type~II breaks} is an issue that has been addressed for a long time and with a renewed vigor during the past couple of years. As we have already referred in Section \ref{sec:intro}, the new rendition of models incorporate the effect of a star formation threshold and the redistribution of stars by secular processes (migration) as key ingredient to match the observational evidence in disk galaxies \citep{Roskar2008,Sanchez-Blazquez2009,Martinez-Serrano2009}. These models have remarkably similar predictions of an existing minimum on the age profile occuring right at the break position followed by a likely ageing towards the outskirts. This was the natural consequence of the net migration of stars. 

The ageing in the outskirts in case of an absent metallicity gradient can be seen as reddening on the color profiles. The Type~II ``U-shape'' (g'-r') color profiles for this reason are in qualitative agreement with the predictions of \citet{Roskar2008,Sanchez-Blazquez2009,Martinez-Serrano2009}. A disagreement between the observational and theoretical results of \citetalias{BTP08} and \citet{Roskar2008} was that \citetalias{BTP08} proposed the \textit{(near) absence of the break} on the stellar surface mass density profile. 

In \citetalias{BTP08} we could not show this phenomena on the individual galaxy profiles due to the insufficient depth of the SDSS data that prevented us to trace the (g'-r') color or the \ssmd\ profiles well beyond the break radius. Although, based on the robust mean \ssmd\ profiles, in \citetalias{BTP08} we claimed that the break on the surface brightness profile signifies a radial change in the key ingreadients of the stellar population rather than being caused by a mass drop beyond the break radius. Based on the results revealed in this paper, we find strong indication that there is no mass drop beyond the break radius. We demonstrated on Figure \ref{fig:break} that the inner-out disk scale-lengths ratio obtained from the stellar surface mass density drops to $\sim$ 1. 

The inner-outer scale-lengths ratio strongly depends on the observing wavelength, as well (see Figure \ref{fig:break}). The prominence of the break decreases from u'-band towards redder filters, in a way that surface brightness profile of the inner disk is the \textit{flattest} in the u'-band (in \citealt{Bakos2010} we observed the same trend on a few selected 'grand-design' spirals.) Consequently, the inner disk \textit{must be} dominated by recently formed, young stars that are distributed homogeneously along the inner disk, meanwhile old stars can be found everywhere in the disk. Remarkably, our \textit{observations} are in qualitative agreement with the predictions of \citet{Sanchez-Blazquez2009} who also showed that the break becomes more pronounced in bluer wavebands. It is however interesting to note, that they find that migration is not the main mechanism producing this particular shape of the age profile. The same trend is observed with and without migration. When looking for explanation for the absence of breaks in the stellar surface mass density profile, however, stellar migration is a key process that weakens the intensity of the break in the \ssmd\ profile. \textit{In situ} star formation in the outer disk cannot produce sufficient amount of stars to conform the lack of or weak breaks. Based on broad-band colors it seems that migration and \textit{in situ} star formation are both relevant. 

On the other hand, we need to stress the importance of the discovery of galaxies like UGC~02081 that based on their surface brightness profile are classified as Type~II but do not exhibit the ``U-shape'', moreover retain a significant mass drop right after the break radius. By finding that the \ssmd\ at the break is at much lower stellar surface mass density ($\sim 1 M_{\odot} pc^{-2}$) than the observed mean value ($\sim 10 M_{\odot} pc^{-2}$), we think we need to treat them as distinct sub-class of Type~II profiles. The low \ssmd\ value at the break and the sharp behavior of the \ssmd\ profile of UGC~02081 at its outskirts resembles more of the traditional picture of truncations (truncations could be imagined as the 'edge' of disks) proposed by \citet{vanderKruit79}. UGC~02081 could be one of the few moderately inclined galaxies where a real truncation is observed.

Finally, in this paper, we propose that Type~III profiles are in fact signatures of the stellar halo (or tidal streams). For this reason, the \textit{formation of Type~III breaks} seems to be a natural consequence of the current cosmological paradigm. Satellites are expected to interact with their host galaxy. Their orbital parameters (angular momentum, for instance) will decide how their interaction changes the morphology of the host galaxy, whether the remnants build the stellar halo or rather settle in the disk \citep{Younger2007,Younger2008,Kazantzidis2008}. We find that in NGC~7716 a disky Type~III morphology can be associated to a misaligned outer disk that is in connection with a tidal stream of an unknown progenitor. The jet-like structure more or less perpendicular to the disk could be part of another loop. These tidal features are well-know around several late-type disks. Just recently, \citet{Wang2012} investigated the possibility that loops in NGC~5907 are formed after a major merger event, challenging the dominance of minor merger scenarios to form such substructures.

\subsection{Stellar halos beyond the local volume}

We observed the presence of stellar halos in 7 spiral galaxies. The signature of the stellar halo is a well-defined upturn followed by a flattening on the deep r'-band surface brightness profiles, that can be observed in the other two deep, g'- and i'-bands.  

Although, the number of systems we can compare to is limited, we can find solid examples of stellar halos observed by 'resolved star' technique in nearby galaxies. If we compare the surface brightness profiles of our galaxies to star-count profiles of Local Group and other nearby galaxies within the next $\sim$15 Mpc, we find that those profiles exhibit the same behavior at similar distances and surface brightness levels. In our galaxies the stellar halo starts to dominate as far as $\sim$ 10 -- 40 kpc, farther in more luminous hosts (see Table \ref{tab:haloprop}). \citet{Courteau2011} reported that in M31 the stellar halo starts to dominate at $\sim$9 kpc over the light of the disk component. NGC2403 (M$_B$=-19.44 mag), a far more isolated system than M31 and analogous to NGC~0941, also exhibit an extended light component that becomes the dominant feature of the stellar count profile at $\sim$18 kpc radius \citep{Barker2012}. 

The light fraction of the stellar halos in our galaxies fall in the range of 1 -- 5 $\%$. These are typical values found in nearby galaxies, the stellar halo light fractions observed in the Milky Way \citep{Carollo2010}, M31 \citep{Courteau2011} or in NGC~2403 \citep{Barker2012} are of the same order we find in our galaxies. The light fraction found in an asymmetric substructure surrounding M33 is $\sim$ 1$\%$ \citep{McConnachie2010}.

According to the findings of \citet{Purcell2007} that investigated the formation of stellar halos through merger-history, the light fraction in stellar halos depends on the total stellar mass of the system and the accretion history of the galaxy, hence on the mass of the dark matter halo. Although, the observed stellar mass (or absolute magnitude) range of our galaxies is of the order of 4 magnitudes, we do not find evidence of the stellar halo light increasing with the host galaxy mass. This might occur if due to the relatively low number of galaxies observed in this sample, the variations in the accretion history of these galaxies \textit{smears} out the correlation expected to be seen with the total galaxy mass. We do find that in the two systems where we discovered the signs of recent interaction the stellar halo light fraction is systematically higher ($\sim$ 10$\%$). 

Other cosmological simulations also emphasize the possibility of a dual origin of stellar halos. In this scenario, stellar halos should contain stars that were formed \textit{in situ} and stars that were accreted from lower mass satellites \citep[e.g.][]{Zolotov2009,Font2011}. Consequently, more massive galaxies are expected to have more metal rich stellar halos. On the observational side there are contradictory findings so far. \citet{Mouhcine2005} suggested that the metallicity of stellar halos in more massive systems is indeed higher, however, \citet{Ferguson2007} showed that the observed photometric metallicities can be largely biased that can be overcome by kinematically selecting the halo star population. As a results of such a careful treatment of halo stars, they found no evidence of correlation between stellar halo metallicity and host galaxy luminosity.

The stellar halo colors we find in our galaxies show no correlation with the host galaxy mass (absolute magnitude) either. We observed a range of different halo colors from red to extremely red, but we have found both 'blue' and red colors in lower-mass systems, namely in NGC~0941 (See Figure \ref{fig:NGC0941_apx}) and UGC~02081 (See Figure \ref{fig:NGC0941_apx}). If not the host galaxy mass then what drives the color distribution of these stellar halos? Do galaxies evolving in isolation have redder colors? Is there any reason to believe that a variant IMF is responsible for the observed colors? Or is there a large population of stars formed \textit{in situ} that could account for the presence of an extremely old but metal-poor population? These are questions that so far remain without answer.

We used the latest MILES SSP models to give crude constraints on the stellar populations embedded within these stellar halos. It is evident that resolved star technique yields more accurate stellar population properties, like metallicities. Broad-brand colors are affected by the age-metallicity degeneracy, so it is not possible to ascertain the stellar populations dominating the stellar halos of our galaxies. The typical stellar halo colors seem be in agreement with the ages and metallicities of nearby galaxies ($\tau \sim$ 5 Gyr and [Fe/H] $\sim$ -1.5). But extreme red halo colors cannot be reconciled with such populations or with the current cosmological paradigm. We find compelling evidence that these extreme red colors may indicate a varying IMF. A bottom-heavy IMF could be the key agent in producing such red populations. Such solution to the red halo phenomenon has also been suggested by several authors \citep[e.g.][]{Zackrisson2006,Zackrisson2012}. 

\section{Conclusions}

\begin{enumerate}

\item The light contribution from the stellar halo in the surface brightness profiles of spiral galaxies causes Type~III profiles: the surface brightness level at which the surface brightness profile starts to deviate from the exponential profile of the disk depends on the brightness of the stellar halo. The brighter the stellar halo, the brighter the Type~III ``break''. Disky Type~III profiles are caused by tidal streams, that also belong to the stellar halo. These effects could be responsible for previous classification of surface brightness profiles of late-type galaxies as Type~III.

\item High surface brightness disk breaks ($\mu_{r'} \sim $22 \sbdim) could be an indication of recent interaction. The new infalling material (gas) settles into the center of the disk and triggers star formation in the inner disk. 

\item The stellar halos of quiescent systems contribute less to the total light of the galaxy.

\item Stellar halos show a large range of ($g'-r'$) colors. Some of them, are extremely red ($g'-r' > $1), allowing for the possibility of bottom-heavy IMFs to explain the red halo colors.

\end{enumerate}

\section{Summary}

In this paper we have explored the structure and stellar population content of 7 late-type spirals using re-stacked deep data from SDSS \str. This data typically reaches $\sim$ 2 magnitudes deeper in the 5 bands than regular SDSS data. In the r'-band the quality of the data made it possible to reach a surface brightness level of $\sim$ 30 mag arcsec$^{-2}$. 

Using the deep r'-band profiles we classified our galaxies into the three classical types (Type~I, Type~II, and Type~III). All surface brightness profiles change their behavior around $\sim$ 28 \sbdim. A flattening is observed on all profiles that we believe to be the signature of stellar halos. 

These faint levels composed of the outer stellar disk and stellar halo beforehand were only reached by techniques that involve resolving stars in these regions or stacking many individual objects. The resolved star technique, although is very powerful, is limited to galaxies that are in the reach of the resolving power of current telescopes. Studies of nearby galaxies like M31, M33, NGC2403 showed that the faint outer regions consist of an ample stellar content with many substructures present in them. We have found different stellar halo color implying very distinct stellar halo populations among our galaxies. We find compelling evidence that there exist extreme red colors halo colors that can be explained by stellar populations that are formed from a bottom-heavy IMF. Meanwhile, less red colors we find in other stellar halos could imply metal-poor ([Fe/H] $\sim$ -1.5) and old ($\tau \gtrsim$ 5 Gyr) stellar populations consistent with the stellar populations found in more nearby galaxies. The light fraction in the stellar halos of our galaxies is 1--5$\%$, but in galaxies where recent interaction occured, it can reach 10$\%$. In NGC~7716 we have discovered the presence of a giant stellar stream, and a $\sim$ -14 r'-mag satellite in the vicinity of NGC~1068. 

\section{Acknowledgement}
This work has been supported by the Programa Nacional de Astronomía y Astrofísica of the Spanish Ministry of Science and Innovation under grant AYA2010-21322-C03-02. JB would like thank the following people who in many ways contributed to this paper: Alexandre Vazdekis, Jes\'us Falc\'on-Barroso, Peter Erwin, Daniella Calzetti, John Kormendy, James Binney, Reynier Peletier, Stephan Courteau and Ignacio Ferreras. 

\bibliography{bakos2012}

\appendix

\section{Galaxy Atlas}\label{sec:galaxy_atlas}

In this section we publish the surface brightness, color, stellar surface mass density profiles of the 7 spiral galaxies of our sample. These figures contain a RGB color composite stamp created from the deep g'-,r'-, and i'-band stacks. These stacks were created using STIFF. In order to help the identification of low surface brightness features by naked eye, we inverted the black background later on. In some cases we also used ``unsharp masking'' to enhance these features. (All images have the same orientation with North displayed upwards and with East to left.)

The top right panel contains the r'-band profile of our galaxies. The plot contains the observed profiles, and the 1$\sigma$(blue shade) and 3$\sigma$(purple shade) error regions. We drew a blue dashed line to indicate the $3\sigma$ noise-level of the sky, the radius where the profile reaches this level is indicated with the blue ellipse on the color image. 

The r'-band profile served as the basis for decomposing the different structures building up the galaxies, like bulge (\textit{green dots}), disk (\textit{black dotted-dashed lines}), and stellar halo (\textit{yellow dots}) components. The decomposition is explained in detail in Section \ref{sec:decomp}. The final composite profile is drawn over the observed profile by \textit{red open circles}. The region where the stellar halo contributes to the total light of the galaxy with more than 50$\%$ is indicated by the green shade. If the surface brightness profile of the galaxy exhibits a break, it is shown by a vertical red dashed line on the profile, and with a red ellipse on the color image. 

The middle left hand panel shows the (g'-r') color profile using scarcely placed bins. For the stellar halo we derived one single color. 

The middle right hand panel shows the \ssmd\ profile (for further information see Section \ref{sec:ssmd}).

The bottom right and left panels contain the surface brightness profiles and color profiles in all available SDSS filters obtained from the \str\ imaging. The surface brightness and color profiles are color coded as indicated on the figures. 

\begin{figure*}[h]
\centering
\includegraphics[width=0.9\textwidth]{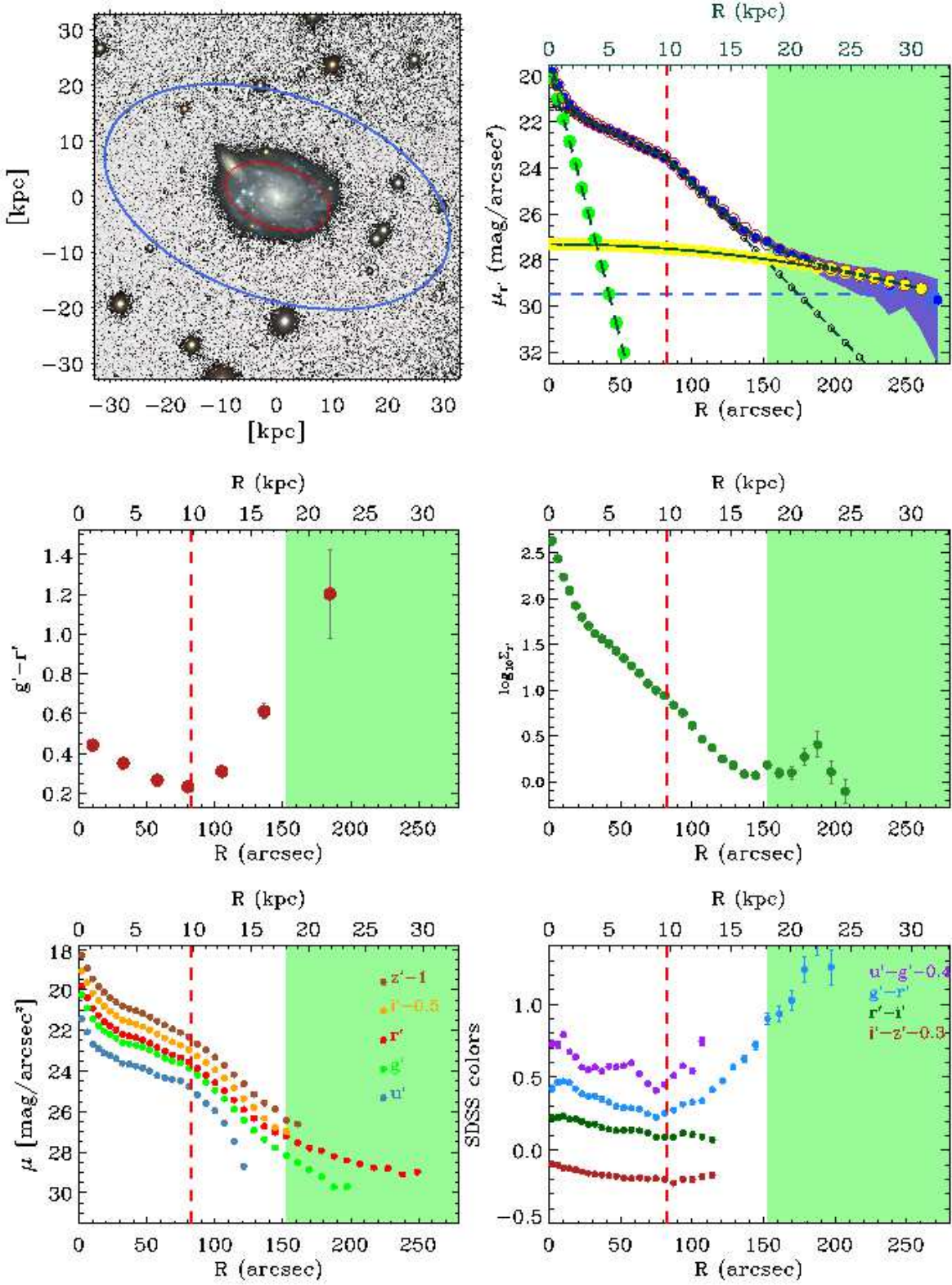}
\caption{NGC~0450: It is a moderately inclined spiral galaxy with a background galaxy (UGC00807) superimposed. It has been classified by \citetalias{PT06} as a Type~II-CT. }
\label{fig:NGC0450_apx}
\end{figure*}

\begin{figure*}[h]
\centering
\includegraphics[width=0.9\textwidth]{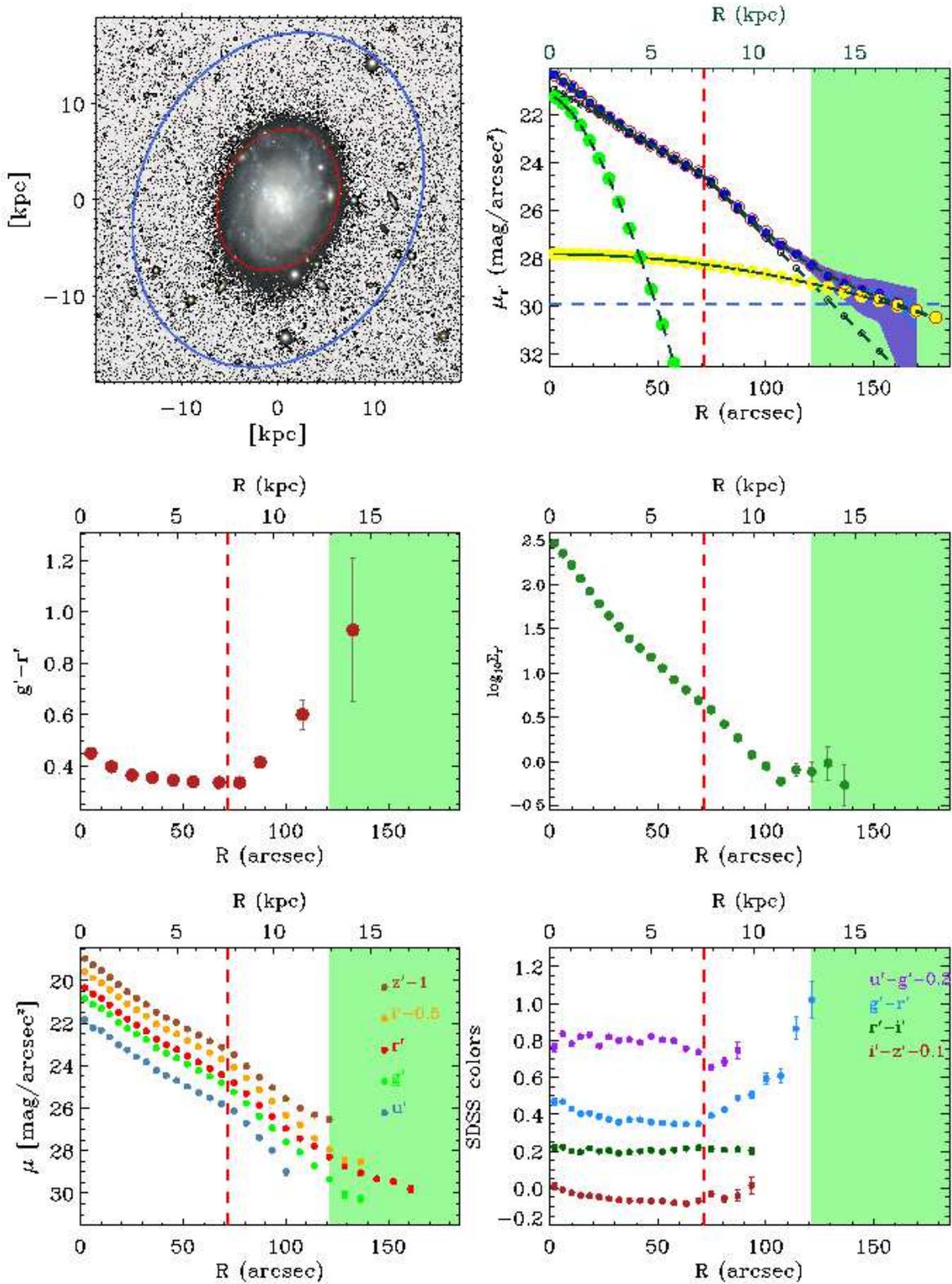}
\caption{NGC~0941: It is a moderately inclined SABc type galaxy. It is within a small group of galaxies, though seems remarkably isolated and the disk shows no signs of interaction. It has a small bar that generates the pseudobulge feature observed at R $\lesssim$ 30$''$ (Kormendy, private communication). The break lies at $\sim$ 70 arcseconds ($\sim$ 7.5 kpc), and is classified as Type~II-CT.}
\label{fig:NGC0941_apx}
\end{figure*}

\begin{figure*}[h]
\centering
\includegraphics[width=0.9\textwidth]{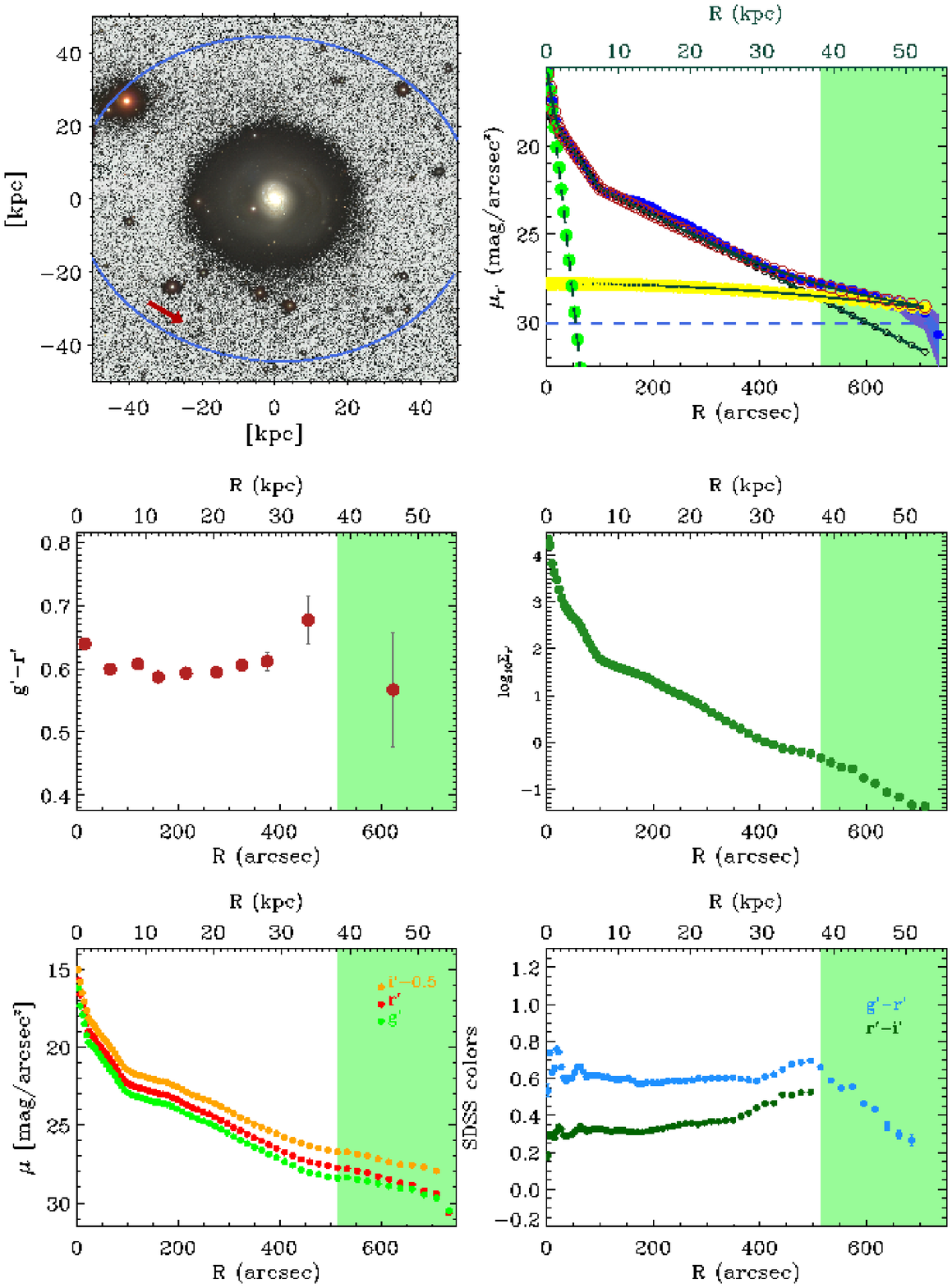}
\caption{NGC~1068 or M77: This galaxy is a frequently studied, bright galaxy, with a Seyfert 2 class AGN \citep{Khachikian1974} in its center. NGC~1068 appears to have a complex inner structure. According to \citet{Erwin2004} this is a double-barred galaxy with a small inner bar (R $<$ 20$''$) embedded within an oval distortion (extending out to R $\lesssim$ 200$''$; \citealt{Schinnerer2000}) also remarked by \citet{Kormendy2004}. We consider the structure outside R $\gtrsim$ 200'' as the genuine disk of the galaxy. This galaxy is basically analogous to NGC~4736 (M94) \citep[see also e.g.][]{Trujillo2009} showing a strikingly similar overall morphology. Considering all these morphological features, this galaxy, based on its r'-band surface brightness profile, has been classified as Type~I. The extended outer disk shows signs of interaction and misalignment, probably a result of recent bombardment. We report here the discovery of a low surface brightness ($\mu_{r'} \sim $27 mag arcsec$^{-2}$) satellite found at the coordinates $\alpha = 02:43:00$ $\delta = -00:09:01$. This object is indicated with a red arrow.}
\label{fig:NGC1068_apx}
\end{figure*}

\begin{figure*}[h]
\centering
\includegraphics[width=0.9\textwidth]{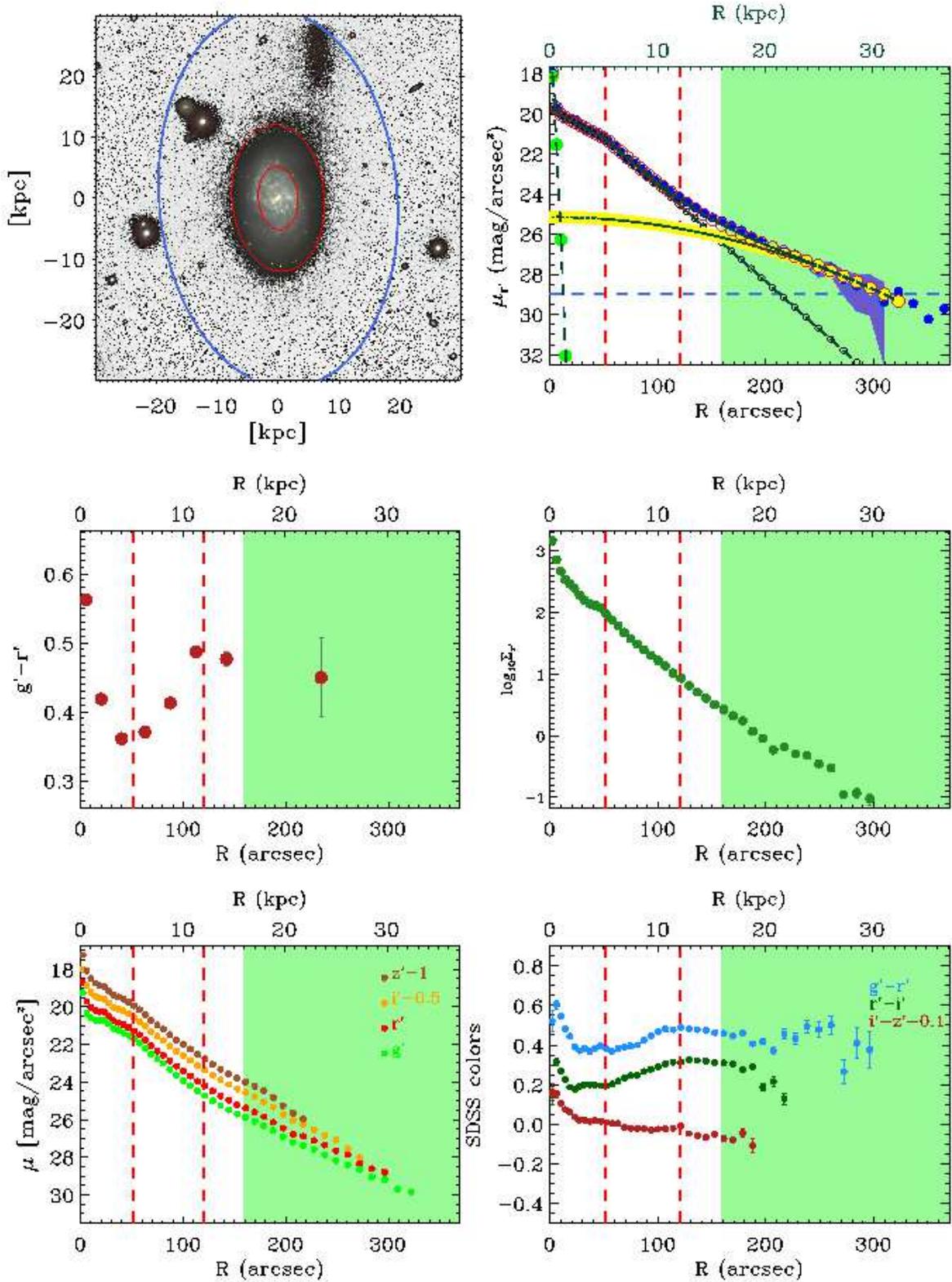}
\caption{NGC~1087: It is a bright, SABc type galaxy with a misaligned outer disk that shows signs of an ongoing interaction on the North side of the galaxy. NGC~1087 is classified as Type II*+III: the apparent Type~II break is due to the central start forming region in the inner disk, meanwhile the bright Type III like feature starting at R $\sim$ 120'' appears because of the presence of an unusually bright stellar halo (due to this, this galaxy previously was classified as Type III in \citetalias{PT06}).}
\label{fig:NGC1087_apx}
\end{figure*}

\begin{figure*}[h]
\centering
\includegraphics[width=0.9\textwidth]{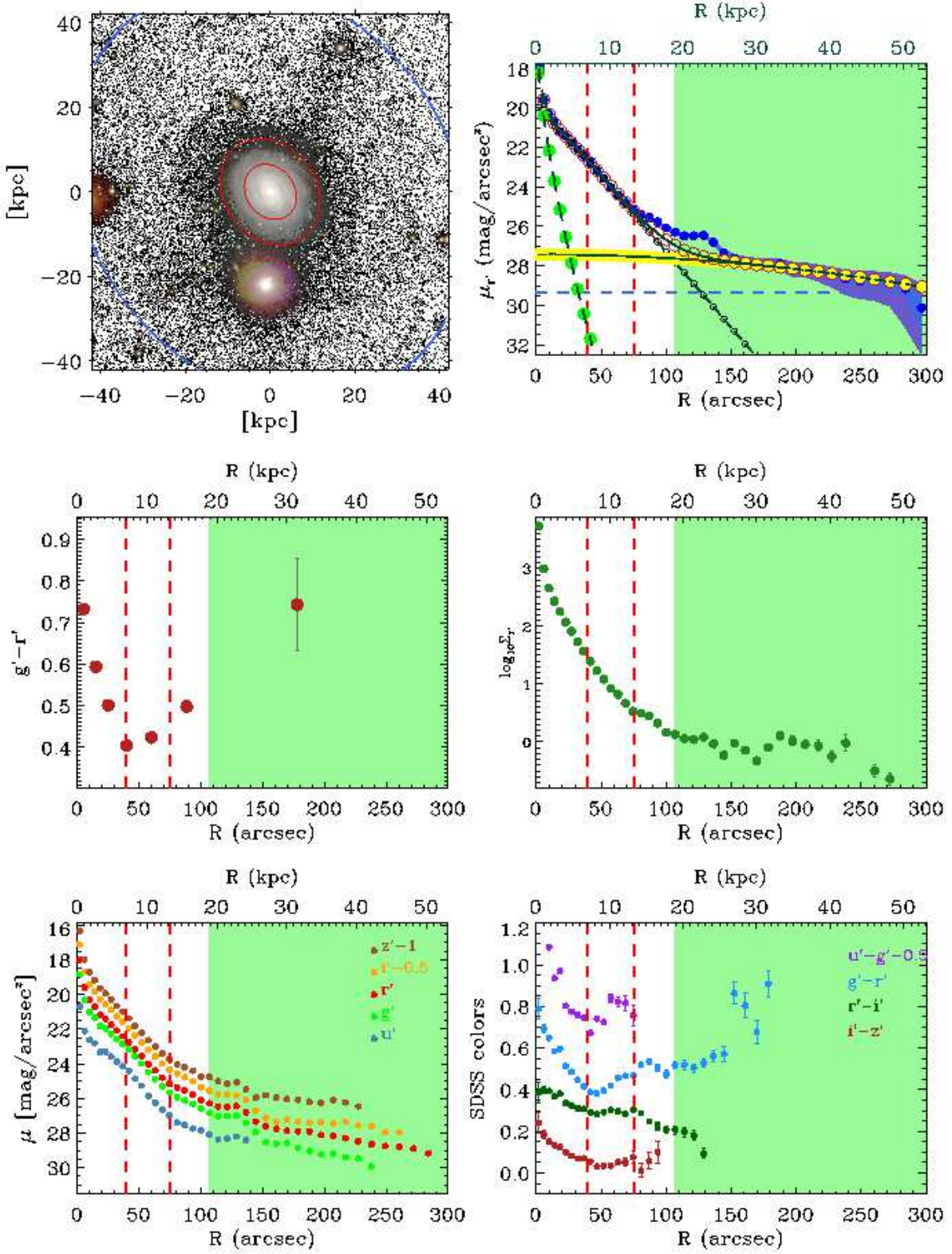}
\caption{NGC~7716: One of the most interesting galaxies of our sample. This galaxy shows no pecularities on its DR7 image, but its \str\ counterpart reveals the presence of a misaligned outer disk and stellar streams. We cannot identify the progenitor satellites of these streams from the images (curiously enough the object that looks like a bright satellite to the North of NGC~7716 is just a background galaxy). Unfortunately, the nearby brights star obscures a part of the stellar stream. The other, jet-like feature is possibly another stream projected along our light-of-sight. NGC~7716 is classified as Type~II*+III, following the previous classification scheme. The Type~III feature, however, is clearly linked to the presence of the bright stellar stream and is part of the stellar halo.}
\label{fig:NGC7716_apx}
\end{figure*}

\begin{figure*}[h]
\centering
\includegraphics[width=0.9\textwidth]{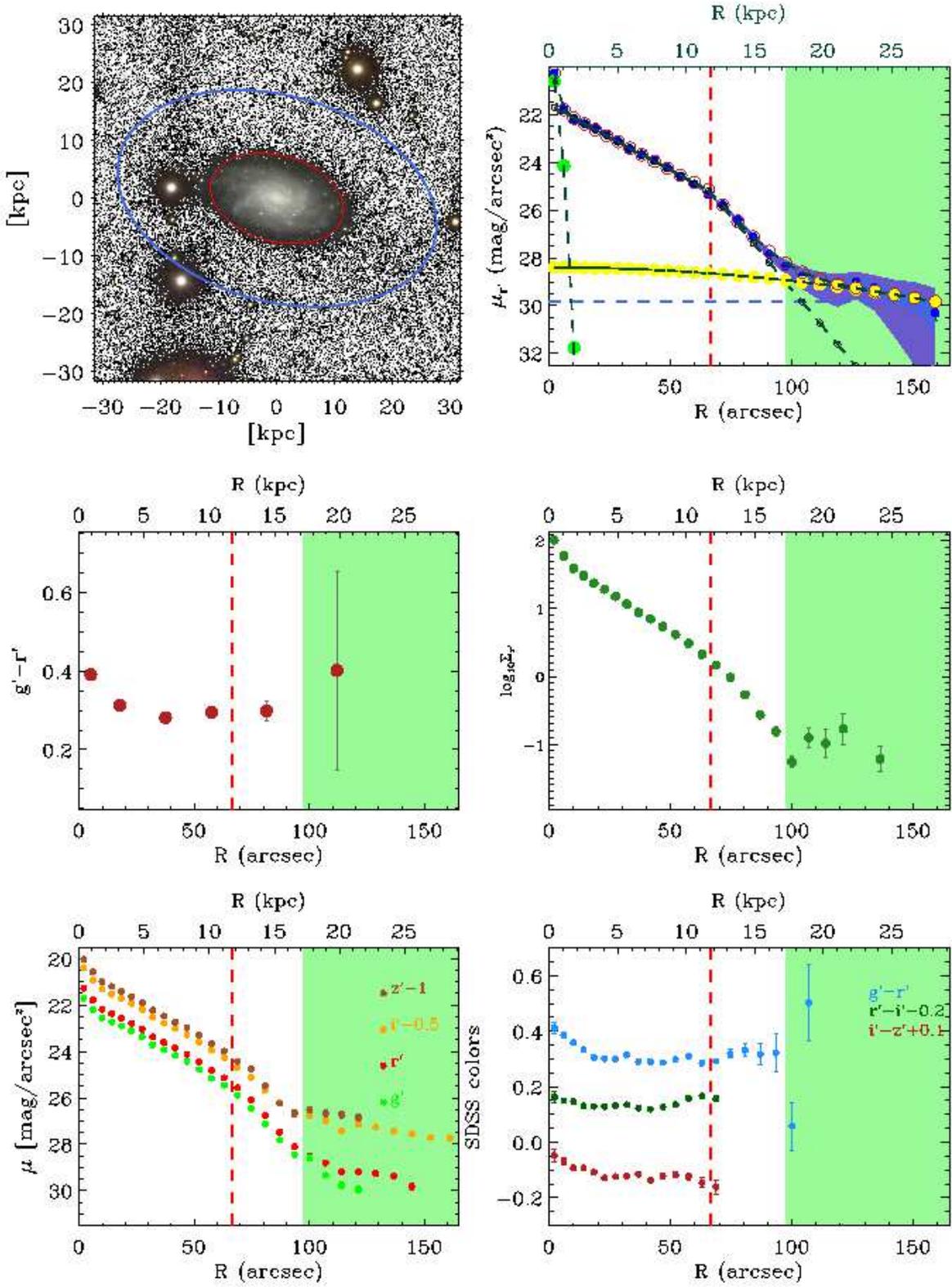}
\caption{UGC~02081: Faint, moderately inclined SABc spiral. The bar is not obvious from the image, nor from the profile. It has a faint break at $\mu_{r'} \sim $ 25 \sbdim\ and is classified as Type~II-CT. }
\label{fig:UGC02081_apx}
\end{figure*}

\begin{figure*}[h]
\centering
\includegraphics[width=0.9\textwidth]{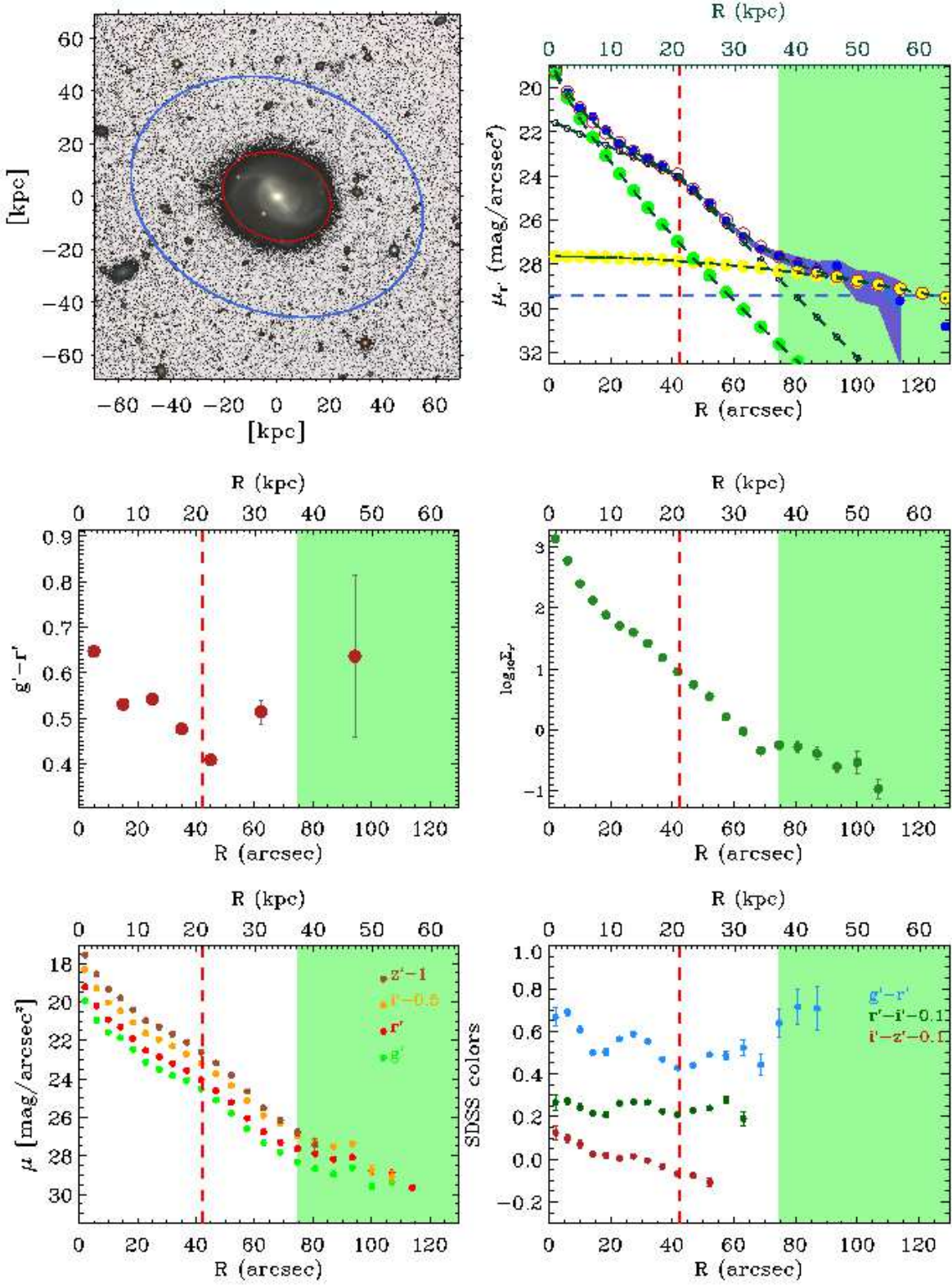}
\caption{UGC~02311: Bright galaxy with a strong bar and with an apparent star forming resonance ring. The break position coincides with the position of the ring, therefore this galaxy is classified as Type~II-ORL.}
\label{fig:UGC02311_apx}
\end{figure*}

\clearpage
\section{Surface Brightness Profiles}
In this section we publish the surface brightness profiles of our galaxies obtained from the \str\ imaging in the 5 SDSS bands. For some of our galaxies, the u'-band and z'-band detector problems prevented us to create reliable mosaics. In these cases (NGC~1068, NGC~1087, UGC~02081, and UGC~02311 in the u'-band and NGC~1068 in the z'-band ) we did not use the u'- and z'-band stacks for further analysis.

\begin{table*}
%
\caption{Surface brightness profiles of NGC0450}
%
\label{tab:NGC0450_sb}
\begin{center}
\renewcommand\tabcolsep{2.00000pt}
\begin{tabular}{cccccc}
\hline
\hline
Radius & $\mu_{u'}$ & $\mu_{g'}$ & $\mu_{r'}$ & $\mu_{i'}$ & $\mu_{z'}$ \\
$$[arcsec] & [mag arcsec$^{-2}$] & [mag arcsec$^{-2}$] & [mag arcsec$^{-2}$] & [mag arcsec$^{-2}$] & [mag arcsec$^{-2}$] \\
\hline
$   2.0$ & $21.44\pm 0.02$ & $20.26\pm 0.02$ & $19.80\pm 0.01$ & $19.55\pm 0.01$ & $19.32\pm 0.01$ \\
\vspace{0.05cm}
$   5.9$ & $22.08\pm 0.02$ & $20.90\pm 0.01$ & $20.40\pm 0.01$ & $20.15\pm 0.01$ & $19.92\pm 0.01$ \\
\vspace{0.05cm}
$  10.0$ & $22.67\pm 0.01$ & $21.43\pm 0.01$ & $20.92\pm 0.01$ & $20.66\pm 0.01$ & $20.45\pm 0.01$ \\
\vspace{0.05cm}
$  14.1$ & $22.90\pm 0.01$ & $21.77\pm 0.01$ & $21.26\pm 0.01$ & $21.03\pm 0.01$ & $20.83\pm 0.01$ \\
\vspace{0.05cm}
$  18.4$ & $23.12\pm 0.01$ & $22.03\pm 0.00$ & $21.56\pm 0.00$ & $21.32\pm 0.00$ & $21.13\pm 0.00$ \\
\vspace{0.05cm}
$  22.8$ & $23.23\pm 0.01$ & $22.20\pm 0.00$ & $21.78\pm 0.00$ & $21.55\pm 0.00$ & $21.38\pm 0.00$ \\
\vspace{0.05cm}
$  27.3$ & $23.40\pm 0.01$ & $22.39\pm 0.01$ & $21.98\pm 0.01$ & $21.77\pm 0.00$ & $21.61\pm 0.01$ \\
\vspace{0.05cm}
$  31.9$ & $23.62\pm 0.01$ & $22.60\pm 0.01$ & $22.18\pm 0.01$ & $21.98\pm 0.00$ & $21.82\pm 0.01$ \\
\vspace{0.05cm}
$  36.7$ & $23.66\pm 0.01$ & $22.66\pm 0.01$ & $22.27\pm 0.01$ & $22.09\pm 0.01$ & $21.93\pm 0.01$ \\
\vspace{0.05cm}
$  41.7$ & $23.76\pm 0.01$ & $22.73\pm 0.01$ & $22.36\pm 0.01$ & $22.19\pm 0.01$ & $22.04\pm 0.01$ \\
\vspace{0.05cm}
$  46.8$ & $23.84\pm 0.01$ & $22.81\pm 0.00$ & $22.47\pm 0.00$ & $22.31\pm 0.00$ & $22.16\pm 0.01$ \\
\vspace{0.05cm}
$  52.0$ & $23.99\pm 0.01$ & $22.96\pm 0.00$ & $22.63\pm 0.00$ & $22.47\pm 0.00$ & $22.33\pm 0.00$ \\
\vspace{0.05cm}
$  57.4$ & $24.22\pm 0.01$ & $23.17\pm 0.00$ & $22.84\pm 0.00$ & $22.67\pm 0.00$ & $22.54\pm 0.00$ \\
\vspace{0.05cm}
$  63.0$ & $24.34\pm 0.01$ & $23.36\pm 0.00$ & $23.03\pm 0.00$ & $22.87\pm 0.00$ & $22.74\pm 0.00$ \\
\vspace{0.05cm}
$  68.7$ & $24.41\pm 0.01$ & $23.51\pm 0.00$ & $23.22\pm 0.00$ & $23.07\pm 0.00$ & $22.94\pm 0.00$ \\
\vspace{0.05cm}
$  74.6$ & $24.47\pm 0.01$ & $23.61\pm 0.00$ & $23.34\pm 0.00$ & $23.22\pm 0.00$ & $23.09\pm 0.01$ \\
\vspace{0.05cm}
$  80.7$ & $24.76\pm 0.01$ & $23.85\pm 0.00$ & $23.56\pm 0.00$ & $23.44\pm 0.00$ & $23.32\pm 0.01$ \\
\vspace{0.05cm}
$  87.0$ & $25.16\pm 0.01$ & $24.19\pm 0.01$ & $23.88\pm 0.00$ & $23.76\pm 0.01$ & $23.66\pm 0.01$ \\
\vspace{0.05cm}
$  93.4$ & $25.59\pm 0.02$ & $24.56\pm 0.01$ & $24.20\pm 0.01$ & $24.06\pm 0.01$ & $23.93\pm 0.01$ \\
\vspace{0.05cm}
$ 100.1$ & $25.95\pm 0.02$ & $24.94\pm 0.01$ & $24.58\pm 0.01$ & $24.44\pm 0.01$ & $24.32\pm 0.01$ \\
\vspace{0.05cm}
$ 106.9$ & $26.56\pm 0.03$ & $25.36\pm 0.01$ & $24.98\pm 0.01$ & $24.86\pm 0.01$ & $24.72\pm 0.01$ \\
\vspace{0.05cm}
$ 113.9$ & $27.46\pm 0.05$ & $25.88\pm 0.01$ & $25.42\pm 0.01$ & $25.33\pm 0.01$ & $25.17\pm 0.02$ \\
\vspace{0.05cm}
$ 121.2$ & $28.70\pm 0.14$ & $26.42\pm 0.01$ & $25.90\pm 0.01$ & $25.86\pm 0.01$ & $25.64\pm 0.03$ \\
\vspace{0.05cm}
$ 128.7$ &  \nodata  & $26.94\pm 0.01$ & $26.33\pm 0.01$ & $26.28\pm 0.02$ & $26.15\pm 0.04$ \\
\vspace{0.05cm}
$ 136.4$ &  \nodata  & $27.38\pm 0.02$ & $26.72\pm 0.01$ & $26.82\pm 0.02$ & $26.57\pm 0.05$ \\
\vspace{0.05cm}
$ 144.3$ &  \nodata  & $27.78\pm 0.03$ & $27.02\pm 0.02$ & $27.27\pm 0.03$ & $26.89\pm 0.07$ \\
\vspace{0.05cm}
$ 152.5$ &  \nodata  & $28.16\pm 0.04$ & $27.22\pm 0.02$ & $27.47\pm 0.04$ & $27.43\pm 0.11$ \\
\vspace{0.05cm}
$ 160.9$ &  \nodata  & $28.51\pm 0.04$ & $27.53\pm 0.02$ & $27.99\pm 0.06$ &  \nodata  \\
\vspace{0.05cm}
$ 169.6$ &  \nodata  & $28.86\pm 0.06$ & $27.79\pm 0.03$ & $28.30\pm 0.07$ &  \nodata  \\
\vspace{0.05cm}
$ 178.5$ &  \nodata  & $29.20\pm 0.08$ & $27.92\pm 0.03$ &  \nodata  &  \nodata  \\
\vspace{0.05cm}
$ 187.7$ &  \nodata  &  \nodata  & $28.20\pm 0.04$ &  \nodata  &  \nodata  \\
\vspace{0.05cm}
$ 197.2$ &  \nodata  &  \nodata  & $28.41\pm 0.04$ &  \nodata  &  \nodata  \\
\vspace{0.05cm}
$ 206.9$ &  \nodata  &  \nodata  & $28.57\pm 0.05$ &  \nodata  &  \nodata  \\
\vspace{0.05cm}
$ 217.0$ &  \nodata  &  \nodata  & $28.76\pm 0.06$ &  \nodata  &  \nodata  \\
\vspace{0.05cm}
$ 227.3$ &  \nodata  &  \nodata  & $28.81\pm 0.06$ &  \nodata  &  \nodata  \\
\vspace{0.05cm}
$ 238.0$ &  \nodata  &  \nodata  & $29.10\pm 0.08$ &  \nodata  &  \nodata  \\
\hline
\hline
\end{tabular}
\end{center}
\end{table*}

\begin{table*}
%
\caption{Surface brightness profiles of NGC0941}
%
\label{tab:NGC0941_sb}
\begin{center}
\renewcommand\tabcolsep{2.00000pt}
\begin{tabular}{cccccc}
\hline
\hline
Radius & $\mu_{u'}$ & $\mu_{g'}$ & $\mu_{r'}$ & $\mu_{i'}$ & $\mu_{z'}$ \\
$$[arcsec] & [mag arcsec$^{-2}$] & [mag arcsec$^{-2}$] & [mag arcsec$^{-2}$] & [mag arcsec$^{-2}$] & [mag arcsec$^{-2}$] \\
\hline
$   2.0$ & $21.84\pm 0.02$ & $20.83\pm 0.02$ & $20.33\pm 0.01$ & $20.08\pm 0.01$ & $19.95\pm 0.01$ \\
\vspace{0.05cm}
$   5.9$ & $22.20\pm 0.01$ & $21.12\pm 0.01$ & $20.62\pm 0.01$ & $20.37\pm 0.00$ & $20.26\pm 0.00$ \\
\vspace{0.05cm}
$  10.0$ & $22.33\pm 0.01$ & $21.30\pm 0.01$ & $20.84\pm 0.01$ & $20.61\pm 0.00$ & $20.51\pm 0.01$ \\
\vspace{0.05cm}
$  14.1$ & $22.65\pm 0.01$ & $21.59\pm 0.01$ & $21.15\pm 0.01$ & $20.93\pm 0.01$ & $20.85\pm 0.01$ \\
\vspace{0.05cm}
$  18.4$ & $23.03\pm 0.01$ & $21.95\pm 0.01$ & $21.51\pm 0.01$ & $21.27\pm 0.00$ & $21.19\pm 0.00$ \\
\vspace{0.05cm}
$  22.8$ & $23.25\pm 0.01$ & $22.24\pm 0.00$ & $21.81\pm 0.00$ & $21.59\pm 0.00$ & $21.51\pm 0.00$ \\
\vspace{0.05cm}
$  27.3$ & $23.58\pm 0.01$ & $22.51\pm 0.00$ & $22.10\pm 0.00$ & $21.88\pm 0.00$ & $21.81\pm 0.00$ \\
\vspace{0.05cm}
$  31.9$ & $23.82\pm 0.01$ & $22.77\pm 0.00$ & $22.38\pm 0.00$ & $22.17\pm 0.00$ & $22.11\pm 0.00$ \\
\vspace{0.05cm}
$  36.7$ & $24.21\pm 0.01$ & $23.16\pm 0.00$ & $22.75\pm 0.00$ & $22.53\pm 0.00$ & $22.48\pm 0.00$ \\
\vspace{0.05cm}
$  41.7$ & $24.45\pm 0.01$ & $23.42\pm 0.00$ & $23.01\pm 0.00$ & $22.79\pm 0.00$ & $22.73\pm 0.00$ \\
\vspace{0.05cm}
$  46.8$ & $24.70\pm 0.01$ & $23.63\pm 0.00$ & $23.24\pm 0.00$ & $23.01\pm 0.00$ & $22.96\pm 0.00$ \\
\vspace{0.05cm}
$  52.0$ & $24.97\pm 0.01$ & $23.92\pm 0.00$ & $23.53\pm 0.00$ & $23.31\pm 0.00$ & $23.26\pm 0.01$ \\
\vspace{0.05cm}
$  57.4$ & $25.27\pm 0.01$ & $24.22\pm 0.00$ & $23.84\pm 0.00$ & $23.61\pm 0.00$ & $23.57\pm 0.01$ \\
\vspace{0.05cm}
$  63.0$ & $25.51\pm 0.01$ & $24.51\pm 0.00$ & $24.12\pm 0.00$ & $23.89\pm 0.00$ & $23.85\pm 0.01$ \\
\vspace{0.05cm}
$  68.7$ & $25.79\pm 0.01$ & $24.81\pm 0.00$ & $24.42\pm 0.00$ & $24.18\pm 0.00$ & $24.13\pm 0.01$ \\
\vspace{0.05cm}
$  74.6$ & $26.14\pm 0.02$ & $25.24\pm 0.01$ & $24.81\pm 0.01$ & $24.57\pm 0.01$ & $24.49\pm 0.01$ \\
\vspace{0.05cm}
$  80.7$ & $26.70\pm 0.02$ & $25.77\pm 0.01$ & $25.31\pm 0.01$ & $25.08\pm 0.01$ & $25.01\pm 0.02$ \\
\vspace{0.05cm}
$  87.0$ & $27.38\pm 0.04$ & $26.38\pm 0.01$ & $25.86\pm 0.01$ & $25.63\pm 0.01$ & $25.54\pm 0.03$ \\
\vspace{0.05cm}
$  93.4$ & $27.99\pm 0.07$ & $26.94\pm 0.01$ & $26.39\pm 0.01$ & $26.17\pm 0.02$ & $26.03\pm 0.04$ \\
\vspace{0.05cm}
$ 100.1$ & $28.99\pm 0.16$ & $27.58\pm 0.02$ & $26.95\pm 0.02$ & $26.81\pm 0.03$ & $26.56\pm 0.06$ \\
\vspace{0.05cm}
$ 106.9$ &  \nodata  & $28.07\pm 0.03$ & $27.43\pm 0.02$ & $27.28\pm 0.03$ & $27.00\pm 0.09$ \\
\vspace{0.05cm}
$ 113.9$ &  \nodata  & $28.70\pm 0.05$ & $27.80\pm 0.03$ & $27.91\pm 0.06$ & $27.31\pm 0.11$ \\
\vspace{0.05cm}
$ 121.2$ &  \nodata  & $29.35\pm 0.09$ & $28.29\pm 0.05$ & $28.44\pm 0.09$ &  \nodata  \\
\vspace{0.05cm}
$ 128.7$ &  \nodata  &  \nodata  & $28.75\pm 0.07$ & $28.93\pm 0.13$ &  \nodata  \\
\vspace{0.05cm}
$ 136.4$ &  \nodata  &  \nodata  & $29.05\pm 0.09$ &  \nodata  &  \nodata  \\
\vspace{0.05cm}
$ 144.3$ &  \nodata  &  \nodata  & $29.33\pm 0.10$ &  \nodata  &  \nodata  \\
\hline
\hline
\end{tabular}
\end{center}
\end{table*}

\begin{table*}
%
\caption{Surface brightness profiles of NGC1068}
%
\label{tab:NGC1068_sb}
\begin{center}
\renewcommand\tabcolsep{2.00000pt}
\renewcommand{\arraystretch}{0.5}
\begin{tabular}{cccc}
\hline
\hline
Radius & $\mu_{g'}$ & $\mu_{r'}$ & $\mu_{i'}$ \\
$$[arcsec] & [mag arcsec$^{-2}$] & [mag arcsec$^{-2}$] & [mag arcsec$^{-2}$] \\
\hline
$   2.0$ & $16.35\pm 0.03$ & $15.78\pm 0.02$ & $15.58\pm 0.02$ \\
\vspace{0.05cm}
$   5.9$ & $17.49\pm 0.01$ & $16.71\pm 0.01$ & $16.40\pm 0.01$ \\
\vspace{0.05cm}
$  10.0$ & $18.06\pm 0.01$ & $17.39\pm 0.01$ & $17.08\pm 0.01$ \\
\vspace{0.05cm}
$  14.1$ & $18.61\pm 0.01$ & $17.92\pm 0.01$ & $17.62\pm 0.01$ \\
\vspace{0.05cm}
$  18.4$ & $19.36\pm 0.01$ & $18.56\pm 0.01$ & $18.23\pm 0.01$ \\
\vspace{0.05cm}
$  22.8$ & $19.84\pm 0.01$ & $19.06\pm 0.01$ & $18.70\pm 0.00$ \\
\vspace{0.05cm}
$  27.3$ & $20.00\pm 0.00$ & $19.30\pm 0.00$ & $18.96\pm 0.00$ \\
\vspace{0.05cm}
$  31.9$ & $20.08\pm 0.00$ & $19.44\pm 0.00$ & $19.13\pm 0.00$ \\
\vspace{0.05cm}
$  36.7$ & $20.27\pm 0.00$ & $19.65\pm 0.00$ & $19.34\pm 0.00$ \\
\vspace{0.05cm}
$  41.7$ & $20.53\pm 0.00$ & $19.90\pm 0.00$ & $19.58\pm 0.00$ \\
\vspace{0.05cm}
$  46.8$ & $20.75\pm 0.00$ & $20.11\pm 0.00$ & $19.79\pm 0.00$ \\
\vspace{0.05cm}
$  52.0$ & $21.00\pm 0.00$ & $20.33\pm 0.00$ & $20.00\pm 0.00$ \\
\vspace{0.05cm}
$  57.4$ & $21.26\pm 0.00$ & $20.57\pm 0.00$ & $20.22\pm 0.00$ \\
\vspace{0.05cm}
$  63.0$ & $21.51\pm 0.00$ & $20.81\pm 0.00$ & $20.45\pm 0.00$ \\
\vspace{0.05cm}
$  68.7$ & $21.72\pm 0.00$ & $21.05\pm 0.00$ & $20.70\pm 0.00$ \\
\vspace{0.05cm}
$  74.6$ & $21.94\pm 0.00$ & $21.30\pm 0.00$ & $20.97\pm 0.00$ \\
\vspace{0.05cm}
$  80.7$ & $22.28\pm 0.00$ & $21.63\pm 0.00$ & $21.29\pm 0.00$ \\
\vspace{0.05cm}
$  87.0$ & $22.60\pm 0.00$ & $21.95\pm 0.00$ & $21.60\pm 0.00$ \\
\vspace{0.05cm}
$  93.4$ & $22.86\pm 0.00$ & $22.21\pm 0.00$ & $21.86\pm 0.00$ \\
\vspace{0.05cm}
$ 100.1$ & $23.06\pm 0.00$ & $22.41\pm 0.00$ & $22.06\pm 0.00$ \\
\vspace{0.05cm}
$ 106.9$ & $23.18\pm 0.00$ & $22.54\pm 0.00$ & $22.20\pm 0.00$ \\
\vspace{0.05cm}
$ 113.9$ & $23.27\pm 0.00$ & $22.62\pm 0.00$ & $22.28\pm 0.00$ \\
\vspace{0.05cm}
$ 121.2$ & $23.32\pm 0.00$ & $22.68\pm 0.00$ & $22.35\pm 0.00$ \\
\vspace{0.05cm}
$ 128.7$ & $23.39\pm 0.00$ & $22.76\pm 0.00$ & $22.42\pm 0.00$ \\
\vspace{0.05cm}
$ 136.4$ & $23.48\pm 0.00$ & $22.85\pm 0.00$ & $22.51\pm 0.00$ \\
\vspace{0.05cm}
$ 144.3$ & $23.57\pm 0.00$ & $22.94\pm 0.00$ & $22.60\pm 0.00$ \\
\vspace{0.05cm}
$ 152.5$ & $23.65\pm 0.00$ & $23.01\pm 0.00$ & $22.68\pm 0.00$ \\
\vspace{0.05cm}
$ 160.9$ & $23.69\pm 0.00$ & $23.06\pm 0.00$ & $22.73\pm 0.00$ \\
\vspace{0.05cm}
$ 169.6$ & $23.72\pm 0.00$ & $23.11\pm 0.00$ & $22.78\pm 0.00$ \\
\vspace{0.05cm}
$ 178.5$ & $23.80\pm 0.00$ & $23.19\pm 0.00$ & $22.86\pm 0.00$ \\
\vspace{0.05cm}
$ 187.7$ & $23.94\pm 0.00$ & $23.33\pm 0.00$ & $22.99\pm 0.00$ \\
\vspace{0.05cm}
$ 197.2$ & $24.07\pm 0.00$ & $23.46\pm 0.00$ & $23.12\pm 0.00$ \\
\vspace{0.05cm}
$ 206.9$ & $24.24\pm 0.00$ & $23.63\pm 0.00$ & $23.29\pm 0.00$ \\
\vspace{0.05cm}
$ 217.0$ & $24.43\pm 0.00$ & $23.81\pm 0.00$ & $23.46\pm 0.00$ \\
\vspace{0.05cm}
$ 227.3$ & $24.60\pm 0.00$ & $23.98\pm 0.00$ & $23.62\pm 0.00$ \\
\vspace{0.05cm}
$ 238.0$ & $24.74\pm 0.00$ & $24.12\pm 0.00$ & $23.75\pm 0.00$ \\
\vspace{0.05cm}
$ 249.0$ & $24.88\pm 0.00$ & $24.26\pm 0.00$ & $23.89\pm 0.00$ \\
\vspace{0.05cm}
$ 260.3$ & $25.01\pm 0.00$ & $24.39\pm 0.00$ & $24.01\pm 0.00$ \\
\vspace{0.05cm}
$ 271.9$ & $25.20\pm 0.00$ & $24.57\pm 0.00$ & $24.19\pm 0.00$ \\
\vspace{0.05cm}
$ 283.9$ & $25.41\pm 0.00$ & $24.77\pm 0.00$ & $24.40\pm 0.00$ \\
\vspace{0.05cm}
$ 296.3$ & $25.62\pm 0.00$ & $24.99\pm 0.00$ & $24.60\pm 0.00$ \\
\vspace{0.05cm}
$ 309.0$ & $25.87\pm 0.00$ & $25.23\pm 0.00$ & $24.84\pm 0.00$ \\
\vspace{0.05cm}
$ 322.1$ & $26.09\pm 0.00$ & $25.46\pm 0.00$ & $25.07\pm 0.00$ \\
\vspace{0.05cm}
$ 335.6$ & $26.32\pm 0.00$ & $25.68\pm 0.00$ & $25.29\pm 0.00$ \\
\vspace{0.05cm}
$ 349.5$ & $26.55\pm 0.00$ & $25.91\pm 0.00$ & $25.51\pm 0.00$ \\
\vspace{0.05cm}
$ 363.9$ & $26.74\pm 0.00$ & $26.11\pm 0.00$ & $25.69\pm 0.00$ \\
\vspace{0.05cm}
$ 378.6$ & $26.99\pm 0.00$ & $26.36\pm 0.00$ & $25.93\pm 0.00$ \\
\vspace{0.05cm}
$ 393.8$ & $27.20\pm 0.00$ & $26.58\pm 0.00$ & $26.12\pm 0.00$ \\
\vspace{0.05cm}
$ 409.5$ & $27.49\pm 0.00$ & $26.84\pm 0.00$ & $26.36\pm 0.00$ \\
\vspace{0.05cm}
$ 425.6$ & $27.69\pm 0.01$ & $27.04\pm 0.00$ & $26.55\pm 0.00$ \\
\vspace{0.05cm}
$ 442.2$ & $27.97\pm 0.01$ & $27.29\pm 0.01$ & $26.75\pm 0.01$ \\
\vspace{0.05cm}
$ 459.3$ & $28.17\pm 0.01$ & $27.46\pm 0.01$ & $26.93\pm 0.01$ \\
\vspace{0.05cm}
$ 476.9$ & $28.31\pm 0.01$ & $27.59\pm 0.01$ & $27.04\pm 0.01$ \\
\vspace{0.05cm}
$ 495.1$ & $28.44\pm 0.01$ & $27.71\pm 0.01$ & $27.17\pm 0.01$ \\
\vspace{0.05cm}
$ 513.8$ & $28.53\pm 0.01$ & $27.83\pm 0.01$ & $27.28\pm 0.01$ \\
\vspace{0.05cm}
$ 533.0$ & $28.50\pm 0.01$ & $27.88\pm 0.01$ & $27.30\pm 0.01$ \\
\vspace{0.05cm}
$ 552.9$ & $28.62\pm 0.01$ & $28.04\pm 0.01$ & $27.44\pm 0.01$ \\
\vspace{0.05cm}
$ 573.3$ & $28.74\pm 0.01$ & $28.15\pm 0.01$ & $27.52\pm 0.01$ \\
\vspace{0.05cm}
$ 594.3$ & $28.89\pm 0.01$ & $28.39\pm 0.01$ & $27.73\pm 0.01$ \\
\vspace{0.05cm}
$ 616.0$ & $29.05\pm 0.01$ & $28.58\pm 0.01$ & $27.94\pm 0.01$ \\
\vspace{0.05cm}
$ 638.3$ & $29.18\pm 0.01$ & $28.81\pm 0.02$ & $28.12\pm 0.01$ \\
\vspace{0.05cm}
$ 661.3$ & $29.25\pm 0.01$ & $28.92\pm 0.02$ & $28.15\pm 0.01$ \\
\vspace{0.05cm}
$ 685.0$ & $29.59\pm 0.02$ & $29.29\pm 0.02$ & $28.27\pm 0.02$ \\
\vspace{0.05cm}
$ 709.4$ & $29.82\pm 0.02$ & $29.47\pm 0.03$ & $28.50\pm 0.02$ \\
\vspace{0.05cm}
$ 734.5$ & $30.64\pm 0.05$ & $30.71\pm 0.08$ &  \nodata  \\
\hline
\hline
\end{tabular}
\end{center}
\end{table*}

\begin{table*}
%
\caption{Surface brightness profiles of NGC1087}
%
\label{tab:NGC1087_sb}
\begin{center}
\renewcommand\tabcolsep{2.00000pt}
\begin{tabular}{ccccc}
\hline
\hline
Radius & $\mu_{g'}$ & $\mu_{r'}$ & $\mu_{i'}$ & $\mu_{z'}$ \\
$$[arcsec] & [mag arcsec$^{-2}$] & [mag arcsec$^{-2}$] & [mag arcsec$^{-2}$] & [mag arcsec$^{-2}$] \\
\hline
$   2.0$ & $19.25\pm 0.04$ & $18.70\pm 0.04$ & $18.53\pm 0.03$ & $18.25\pm 0.03$ \\
\vspace{0.05cm}
$   5.9$ & $20.35\pm 0.01$ & $19.71\pm 0.01$ & $19.37\pm 0.01$ & $19.10\pm 0.01$ \\
\vspace{0.05cm}
$  10.0$ & $20.60\pm 0.01$ & $20.02\pm 0.01$ & $19.73\pm 0.01$ & $19.50\pm 0.01$ \\
\vspace{0.05cm}
$  14.1$ & $20.70\pm 0.01$ & $20.18\pm 0.01$ & $19.94\pm 0.00$ & $19.74\pm 0.01$ \\
\vspace{0.05cm}
$  18.4$ & $20.74\pm 0.00$ & $20.26\pm 0.00$ & $20.05\pm 0.00$ & $19.87\pm 0.00$ \\
\vspace{0.05cm}
$  22.8$ & $20.71\pm 0.00$ & $20.28\pm 0.00$ & $20.08\pm 0.00$ & $19.92\pm 0.00$ \\
\vspace{0.05cm}
$  27.3$ & $20.91\pm 0.01$ & $20.51\pm 0.00$ & $20.29\pm 0.00$ & $20.15\pm 0.00$ \\
\vspace{0.05cm}
$  31.9$ & $21.16\pm 0.00$ & $20.74\pm 0.00$ & $20.52\pm 0.00$ & $20.38\pm 0.00$ \\
\vspace{0.05cm}
$  36.7$ & $21.24\pm 0.00$ & $20.84\pm 0.00$ & $20.62\pm 0.00$ & $20.48\pm 0.00$ \\
\vspace{0.05cm}
$  41.7$ & $21.37\pm 0.00$ & $20.95\pm 0.00$ & $20.73\pm 0.00$ & $20.59\pm 0.00$ \\
\vspace{0.05cm}
$  46.8$ & $21.52\pm 0.00$ & $21.09\pm 0.00$ & $20.87\pm 0.00$ & $20.74\pm 0.00$ \\
\vspace{0.05cm}
$  52.0$ & $21.71\pm 0.00$ & $21.29\pm 0.00$ & $21.08\pm 0.00$ & $20.94\pm 0.00$ \\
\vspace{0.05cm}
$  57.4$ & $21.91\pm 0.00$ & $21.50\pm 0.00$ & $21.28\pm 0.00$ & $21.15\pm 0.00$ \\
\vspace{0.05cm}
$  63.0$ & $22.20\pm 0.00$ & $21.78\pm 0.00$ & $21.54\pm 0.00$ & $21.42\pm 0.00$ \\
\vspace{0.05cm}
$  68.7$ & $22.50\pm 0.00$ & $22.08\pm 0.00$ & $21.82\pm 0.00$ & $21.72\pm 0.00$ \\
\vspace{0.05cm}
$  74.6$ & $22.78\pm 0.00$ & $22.34\pm 0.00$ & $22.08\pm 0.00$ & $21.98\pm 0.00$ \\
\vspace{0.05cm}
$  80.7$ & $23.01\pm 0.00$ & $22.57\pm 0.00$ & $22.30\pm 0.00$ & $22.20\pm 0.00$ \\
\vspace{0.05cm}
$  87.0$ & $23.33\pm 0.00$ & $22.87\pm 0.00$ & $22.58\pm 0.00$ & $22.48\pm 0.00$ \\
\vspace{0.05cm}
$  93.4$ & $23.64\pm 0.00$ & $23.16\pm 0.00$ & $22.86\pm 0.00$ & $22.76\pm 0.00$ \\
\vspace{0.05cm}
$ 100.1$ & $23.93\pm 0.00$ & $23.42\pm 0.00$ & $23.11\pm 0.00$ & $23.01\pm 0.00$ \\
\vspace{0.05cm}
$ 106.9$ & $24.20\pm 0.00$ & $23.68\pm 0.00$ & $23.35\pm 0.00$ & $23.26\pm 0.00$ \\
\vspace{0.05cm}
$ 113.9$ & $24.48\pm 0.00$ & $23.97\pm 0.00$ & $23.63\pm 0.00$ & $23.53\pm 0.00$ \\
\vspace{0.05cm}
$ 121.2$ & $24.73\pm 0.00$ & $24.20\pm 0.00$ & $23.87\pm 0.00$ & $23.75\pm 0.01$ \\
\vspace{0.05cm}
$ 128.7$ & $25.01\pm 0.00$ & $24.49\pm 0.00$ & $24.14\pm 0.00$ & $24.07\pm 0.01$ \\
\vspace{0.05cm}
$ 136.4$ & $25.25\pm 0.00$ & $24.74\pm 0.00$ & $24.39\pm 0.00$ & $24.33\pm 0.01$ \\
\vspace{0.05cm}
$ 144.3$ & $25.48\pm 0.00$ & $24.97\pm 0.00$ & $24.63\pm 0.00$ & $24.57\pm 0.01$ \\
\vspace{0.05cm}
$ 152.5$ & $25.70\pm 0.00$ & $25.20\pm 0.00$ & $24.86\pm 0.00$ & $24.80\pm 0.01$ \\
\vspace{0.05cm}
$ 160.9$ & $25.88\pm 0.00$ & $25.39\pm 0.00$ & $25.05\pm 0.00$ & $25.00\pm 0.01$ \\
\vspace{0.05cm}
$ 169.6$ & $26.09\pm 0.00$ & $25.60\pm 0.00$ & $25.28\pm 0.01$ & $25.23\pm 0.02$ \\
\vspace{0.05cm}
$ 178.5$ & $26.35\pm 0.01$ & $25.85\pm 0.01$ & $25.55\pm 0.01$ & $25.47\pm 0.02$ \\
\vspace{0.05cm}
$ 187.7$ & $26.58\pm 0.01$ & $26.14\pm 0.01$ & $25.83\pm 0.01$ & $25.82\pm 0.03$ \\
\vspace{0.05cm}
$ 197.2$ & $26.90\pm 0.01$ & $26.45\pm 0.01$ & $26.24\pm 0.01$ & $26.26\pm 0.05$ \\
\vspace{0.05cm}
$ 206.9$ & $27.20\pm 0.01$ & $26.79\pm 0.02$ & $26.55\pm 0.02$ & $26.65\pm 0.07$ \\
\vspace{0.05cm}
$ 217.0$ & $27.38\pm 0.02$ & $26.89\pm 0.02$ & $26.74\pm 0.02$ & $26.98\pm 0.09$ \\
\vspace{0.05cm}
$ 227.3$ & $27.58\pm 0.02$ & $27.11\pm 0.02$ & $27.05\pm 0.03$ & $27.26\pm 0.11$ \\
\vspace{0.05cm}
$ 238.0$ & $27.88\pm 0.02$ & $27.35\pm 0.02$ & $27.35\pm 0.04$ & $27.98\pm 0.20$ \\
\vspace{0.05cm}
$ 249.0$ & $28.17\pm 0.03$ & $27.66\pm 0.03$ & $27.57\pm 0.04$ & $28.08\pm 0.20$ \\
\vspace{0.05cm}
$ 260.3$ & $28.40\pm 0.03$ & $27.87\pm 0.03$ & $28.05\pm 0.06$ &  \nodata  \\
\vspace{0.05cm}
$ 271.9$ & $28.63\pm 0.04$ & $28.33\pm 0.05$ & $28.51\pm 0.08$ &  \nodata  \\
\vspace{0.05cm}
$ 283.9$ & $29.05\pm 0.05$ & $28.61\pm 0.06$ & $29.04\pm 0.13$ &  \nodata  \\
\vspace{0.05cm}
$ 296.3$ & $29.19\pm 0.06$ & $28.78\pm 0.07$ & $30.47\pm 0.42$ &  \nodata  \\
\vspace{0.05cm}
$ 309.0$ & $29.70\pm 0.10$ & $29.38\pm 0.12$ &  \nodata  &  \nodata  \\
\vspace{0.05cm}
$ 322.1$ & $29.85\pm 0.11$ & $29.04\pm 0.09$ &  \nodata  &  \nodata  \\
\vspace{0.05cm}
$ 335.6$ & $30.44\pm 0.17$ & $29.33\pm 0.11$ &  \nodata  &  \nodata  \\
\vspace{0.05cm}
$ 349.5$ &  \nodata  & $31.36\pm 0.55$ &  \nodata  &  \nodata  \\
\hline
\hline
\end{tabular}
\end{center}
\end{table*}

\begin{table*}
%
\caption{Surface brightness profiles of NGC7716}
%
\label{tab:NGC7716_sb}
\begin{center}
\renewcommand\tabcolsep{2.00000pt}
\begin{tabular}{cccccc}
\hline
\hline
Radius & $\mu_{u'}$ & $\mu_{g'}$ & $\mu_{r'}$ & $\mu_{i'}$ & $\mu_{z'}$ \\
$$[arcsec] & [mag arcsec$^{-2}$] & [mag arcsec$^{-2}$] & [mag arcsec$^{-2}$] & [mag arcsec$^{-2}$] & [mag arcsec$^{-2}$] \\
\hline
$   2.0$ & $20.69\pm 0.03$ & $18.84\pm 0.03$ & $18.01\pm 0.04$ & $17.60\pm 0.04$ & $17.34\pm 0.04$ \\
\vspace{0.05cm}
$   5.9$ & $22.11\pm 0.01$ & $20.32\pm 0.01$ & $19.60\pm 0.01$ & $19.17\pm 0.01$ & $18.97\pm 0.01$ \\
\vspace{0.05cm}
$  10.0$ & $22.63\pm 0.00$ & $21.00\pm 0.00$ & $20.32\pm 0.00$ & $19.90\pm 0.00$ & $19.73\pm 0.00$ \\
\vspace{0.05cm}
$  14.1$ & $22.81\pm 0.01$ & $21.33\pm 0.01$ & $20.71\pm 0.01$ & $20.32\pm 0.00$ & $20.16\pm 0.00$ \\
\vspace{0.05cm}
$  18.4$ & $23.33\pm 0.01$ & $21.82\pm 0.01$ & $21.18\pm 0.00$ & $20.78\pm 0.00$ & $20.64\pm 0.00$ \\
\vspace{0.05cm}
$  22.8$ & $23.31\pm 0.01$ & $21.96\pm 0.00$ & $21.42\pm 0.00$ & $21.06\pm 0.00$ & $20.94\pm 0.00$ \\
\vspace{0.05cm}
$  27.3$ & $23.59\pm 0.01$ & $22.26\pm 0.00$ & $21.74\pm 0.00$ & $21.40\pm 0.00$ & $21.30\pm 0.00$ \\
\vspace{0.05cm}
$  31.9$ & $23.87\pm 0.01$ & $22.57\pm 0.00$ & $22.09\pm 0.00$ & $21.75\pm 0.00$ & $21.67\pm 0.00$ \\
\vspace{0.05cm}
$  36.7$ & $24.15\pm 0.01$ & $22.86\pm 0.00$ & $22.41\pm 0.00$ & $22.08\pm 0.00$ & $21.99\pm 0.00$ \\
\vspace{0.05cm}
$  41.7$ & $24.41\pm 0.01$ & $23.19\pm 0.00$ & $22.77\pm 0.00$ & $22.45\pm 0.00$ & $22.38\pm 0.00$ \\
\vspace{0.05cm}
$  46.8$ & $24.87\pm 0.01$ & $23.58\pm 0.00$ & $23.16\pm 0.00$ & $22.86\pm 0.00$ & $22.80\pm 0.01$ \\
\vspace{0.05cm}
$  52.0$ & $25.26\pm 0.01$ & $23.99\pm 0.00$ & $23.56\pm 0.00$ & $23.25\pm 0.00$ & $23.20\pm 0.01$ \\
\vspace{0.05cm}
$  57.4$ & $25.86\pm 0.02$ & $24.48\pm 0.00$ & $24.02\pm 0.00$ & $23.70\pm 0.00$ & $23.64\pm 0.01$ \\
\vspace{0.05cm}
$  63.0$ & $26.25\pm 0.02$ & $24.88\pm 0.00$ & $24.39\pm 0.00$ & $24.07\pm 0.00$ & $24.00\pm 0.01$ \\
\vspace{0.05cm}
$  68.7$ & $26.66\pm 0.04$ & $25.30\pm 0.01$ & $24.80\pm 0.01$ & $24.49\pm 0.01$ & $24.42\pm 0.02$ \\
\vspace{0.05cm}
$  74.6$ & $26.97\pm 0.05$ & $25.67\pm 0.01$ & $25.16\pm 0.01$ & $24.83\pm 0.01$ & $24.74\pm 0.03$ \\
\vspace{0.05cm}
$  80.7$ & $27.38\pm 0.06$ & $25.93\pm 0.01$ & $25.38\pm 0.01$ & $25.07\pm 0.01$ & $25.04\pm 0.03$ \\
\vspace{0.05cm}
$  87.0$ & $27.55\pm 0.08$ & $26.12\pm 0.01$ & $25.55\pm 0.01$ & $25.28\pm 0.01$ & $25.20\pm 0.04$ \\
\vspace{0.05cm}
$  93.4$ & $27.73\pm 0.09$ & $26.31\pm 0.01$ & $25.77\pm 0.01$ & $25.52\pm 0.01$ & $25.40\pm 0.05$ \\
\vspace{0.05cm}
$ 100.1$ & $27.84\pm 0.09$ & $26.59\pm 0.01$ & $26.08\pm 0.01$ & $25.85\pm 0.02$ & $25.73\pm 0.06$ \\
\vspace{0.05cm}
$ 106.9$ & $28.09\pm 0.10$ & $26.85\pm 0.01$ & $26.29\pm 0.01$ & $26.06\pm 0.02$ & $25.77\pm 0.05$ \\
\vspace{0.05cm}
$ 113.9$ & $28.36\pm 0.12$ & $27.03\pm 0.02$ & $26.47\pm 0.01$ & $26.26\pm 0.02$ & $26.02\pm 0.06$ \\
\vspace{0.05cm}
$ 121.2$ & $28.33\pm 0.11$ & $27.01\pm 0.01$ & $26.47\pm 0.01$ & $26.27\pm 0.02$ & $26.16\pm 0.07$ \\
\vspace{0.05cm}
$ 128.7$ & $28.20\pm 0.10$ & $27.01\pm 0.01$ & $26.44\pm 0.01$ & $26.33\pm 0.02$ & $26.06\pm 0.06$ \\
\vspace{0.05cm}
$ 136.4$ & $28.41\pm 0.11$ & $27.40\pm 0.02$ & $26.81\pm 0.02$ & $26.58\pm 0.02$ & $26.47\pm 0.08$ \\
\vspace{0.05cm}
$ 144.3$ & $29.24\pm 0.20$ & $27.94\pm 0.03$ & $27.34\pm 0.02$ & $27.20\pm 0.03$ & $26.95\pm 0.11$ \\
\vspace{0.05cm}
$ 152.5$ & $29.11\pm 0.18$ & $28.53\pm 0.04$ & $27.63\pm 0.03$ & $27.61\pm 0.04$ & $26.82\pm 0.09$ \\
\vspace{0.05cm}
$ 160.9$ & $29.27\pm 0.20$ & $28.60\pm 0.05$ & $27.76\pm 0.03$ & $27.86\pm 0.06$ & $26.95\pm 0.10$ \\
\vspace{0.05cm}
$ 169.6$ & $29.49\pm 0.22$ & $28.57\pm 0.04$ & $27.86\pm 0.03$ & $27.72\pm 0.05$ & $27.05\pm 0.11$ \\
\vspace{0.05cm}
$ 178.5$ &  \nodata  & $28.85\pm 0.05$ & $27.91\pm 0.03$ & $27.86\pm 0.05$ & $27.07\pm 0.10$ \\
\vspace{0.05cm}
$ 187.7$ &  \nodata  & $29.03\pm 0.06$ & $27.91\pm 0.03$ & $27.90\pm 0.05$ & $27.21\pm 0.11$ \\
\vspace{0.05cm}
$ 197.2$ &  \nodata  & $29.22\pm 0.07$ & $28.11\pm 0.04$ & $27.87\pm 0.05$ & $27.25\pm 0.11$ \\
\vspace{0.05cm}
$ 206.9$ &  \nodata  & $29.21\pm 0.06$ & $28.14\pm 0.04$ & $27.95\pm 0.05$ & $27.09\pm 0.10$ \\
\vspace{0.05cm}
$ 217.0$ &  \nodata  & $29.39\pm 0.07$ & $28.29\pm 0.04$ & $27.84\pm 0.04$ & $27.23\pm 0.10$ \\
\vspace{0.05cm}
$ 227.3$ &  \nodata  & $29.47\pm 0.08$ & $28.46\pm 0.04$ & $28.06\pm 0.05$ & $27.46\pm 0.13$ \\
\vspace{0.05cm}
$ 238.0$ &  \nodata  & $29.94\pm 0.12$ & $28.66\pm 0.06$ & $28.10\pm 0.05$ & $27.56\pm 0.14$ \\
\vspace{0.05cm}
$ 249.0$ &  \nodata  & $30.41\pm 0.18$ & $28.75\pm 0.06$ & $28.46\pm 0.07$ & $27.42\pm 0.12$ \\
\vspace{0.05cm}
$ 260.3$ &  \nodata  & $29.68\pm 0.09$ & $28.79\pm 0.06$ & $28.46\pm 0.07$ & $27.77\pm 0.16$ \\
\vspace{0.05cm}
$ 271.9$ &  \nodata  & $29.69\pm 0.09$ & $28.90\pm 0.06$ & $28.55\pm 0.07$ & $27.57\pm 0.13$ \\
\vspace{0.05cm}
$ 283.9$ &  \nodata  & $30.13\pm 0.12$ & $29.16\pm 0.08$ & $28.98\pm 0.10$ & $27.88\pm 0.17$ \\
\vspace{0.05cm}
$ 296.3$ &  \nodata  & $32.10\pm 0.56$ & $30.14\pm 0.16$ & $30.59\pm 0.36$ & $27.98\pm 0.17$ \\
\vspace{0.05cm}
$ 309.0$ &  \nodata  &  \nodata  & $29.99\pm 0.14$ &  \nodata  & $28.93\pm 0.36$ \\
\vspace{0.05cm}
$ 322.1$ &  \nodata  &  \nodata  & $29.90\pm 0.12$ &  \nodata  &  \nodata  \\
\vspace{0.05cm}
$ 335.6$ &  \nodata  &  \nodata  & $30.17\pm 0.15$ &  \nodata  &  \nodata  \\
\vspace{0.05cm}
$ 349.5$ &  \nodata  &  \nodata  & $31.03\pm 0.30$ &  \nodata  &  \nodata  \\
\hline
\hline
\end{tabular}
\end{center}
\end{table*}

\begin{table*}
%
\caption{Surface brightness profiles of UGC02081}
%
\label{tab:UGC02081_sb}
\begin{center}
\renewcommand\tabcolsep{2.00000pt}
\begin{tabular}{ccccc}
\hline
\hline
Radius & $\mu_{g'}$ & $\mu_{r'}$ & $\mu_{i'}$ & $\mu_{z'}$ \\
$$[arcsec] & [mag arcsec$^{-2}$] & [mag arcsec$^{-2}$] & [mag arcsec$^{-2}$] & [mag arcsec$^{-2}$] \\
\hline
$   2.0$ & $21.71\pm 0.01$ & $21.27\pm 0.01$ & $20.89\pm 0.02$ & $21.02\pm 0.02$ \\
\vspace{0.05cm}
$   5.9$ & $22.21\pm 0.01$ & $21.79\pm 0.01$ & $21.42\pm 0.01$ & $21.58\pm 0.01$ \\
\vspace{0.05cm}
$  10.0$ & $22.56\pm 0.00$ & $22.18\pm 0.00$ & $21.81\pm 0.00$ & $21.99\pm 0.00$ \\
\vspace{0.05cm}
$  14.1$ & $22.73\pm 0.00$ & $22.37\pm 0.00$ & $22.02\pm 0.00$ & $22.19\pm 0.00$ \\
\vspace{0.05cm}
$  18.4$ & $22.90\pm 0.00$ & $22.57\pm 0.00$ & $22.22\pm 0.00$ & $22.42\pm 0.00$ \\
\vspace{0.05cm}
$  22.8$ & $23.11\pm 0.00$ & $22.78\pm 0.00$ & $22.44\pm 0.00$ & $22.65\pm 0.00$ \\
\vspace{0.05cm}
$  27.3$ & $23.37\pm 0.00$ & $23.04\pm 0.00$ & $22.69\pm 0.00$ & $22.90\pm 0.00$ \\
\vspace{0.05cm}
$  31.9$ & $23.71\pm 0.00$ & $23.37\pm 0.00$ & $23.02\pm 0.00$ & $23.22\pm 0.01$ \\
\vspace{0.05cm}
$  36.7$ & $23.93\pm 0.00$ & $23.61\pm 0.00$ & $23.27\pm 0.00$ & $23.47\pm 0.01$ \\
\vspace{0.05cm}
$  41.7$ & $24.17\pm 0.00$ & $23.85\pm 0.00$ & $23.51\pm 0.00$ & $23.73\pm 0.01$ \\
\vspace{0.05cm}
$  46.8$ & $24.43\pm 0.00$ & $24.12\pm 0.00$ & $23.77\pm 0.00$ & $23.98\pm 0.01$ \\
\vspace{0.05cm}
$  52.0$ & $24.77\pm 0.00$ & $24.45\pm 0.00$ & $24.10\pm 0.00$ & $24.30\pm 0.01$ \\
\vspace{0.05cm}
$  57.4$ & $25.15\pm 0.00$ & $24.81\pm 0.00$ & $24.43\pm 0.00$ & $24.64\pm 0.01$ \\
\vspace{0.05cm}
$  63.0$ & $25.45\pm 0.00$ & $25.14\pm 0.00$ & $24.76\pm 0.01$ & $24.99\pm 0.02$ \\
\vspace{0.05cm}
$  68.7$ & $25.88\pm 0.01$ & $25.56\pm 0.01$ & $25.19\pm 0.01$ & $25.44\pm 0.02$ \\
\vspace{0.05cm}
$  74.6$ & $26.44\pm 0.01$ & $26.09\pm 0.01$ & $25.63\pm 0.01$ & $25.76\pm 0.03$ \\
\vspace{0.05cm}
$  80.7$ & $27.12\pm 0.02$ & $26.76\pm 0.02$ & $26.20\pm 0.02$ & $26.48\pm 0.06$ \\
\vspace{0.05cm}
$  87.0$ & $27.81\pm 0.03$ & $27.47\pm 0.03$ & $26.70\pm 0.02$ & $27.24\pm 0.11$ \\
\vspace{0.05cm}
$  93.4$ & $28.45\pm 0.05$ & $28.10\pm 0.05$ & $27.13\pm 0.04$ & $27.66\pm 0.16$ \\
\vspace{0.05cm}
$ 100.1$ & $28.58\pm 0.05$ & $28.49\pm 0.07$ & $27.26\pm 0.04$ & $27.51\pm 0.14$ \\
\vspace{0.05cm}
$ 106.9$ & $29.34\pm 0.10$ & $28.80\pm 0.09$ & $27.49\pm 0.05$ & $27.67\pm 0.16$ \\
\vspace{0.05cm}
$ 113.9$ & $29.78\pm 0.14$ & $29.19\pm 0.12$ & $27.92\pm 0.07$ & $27.71\pm 0.16$ \\
\vspace{0.05cm}
$ 121.2$ & $29.94\pm 0.16$ & $29.17\pm 0.11$ & $27.65\pm 0.05$ & $27.85\pm 0.18$ \\
\vspace{0.05cm}
$ 128.7$ & $30.59\pm 0.25$ & $29.26\pm 0.10$ & $27.77\pm 0.05$ & $28.21\pm 0.21$ \\
\vspace{0.05cm}
$ 136.4$ &  \nodata  & $29.37\pm 0.12$ & $27.92\pm 0.06$ &  \nodata  \\
\vspace{0.05cm}
$ 144.3$ &  \nodata  & $29.82\pm 0.16$ & $28.05\pm 0.06$ &  \nodata  \\
\vspace{0.05cm}
$ 152.5$ &  \nodata  & $29.81\pm 0.15$ & $28.20\pm 0.07$ &  \nodata  \\
\vspace{0.05cm}
$ 160.9$ &  \nodata  & $30.60\pm 0.29$ & $28.24\pm 0.07$ &  \nodata  \\
\vspace{0.05cm}
$ 169.6$ &  \nodata  &  \nodata  & $28.47\pm 0.08$ &  \nodata  \\
\vspace{0.05cm}
$ 178.5$ &  \nodata  &  \nodata  & $28.66\pm 0.10$ &  \nodata  \\
\vspace{0.05cm}
$ 187.7$ &  \nodata  &  \nodata  & $29.00\pm 0.13$ &  \nodata  \\
\vspace{0.05cm}
$ 197.2$ &  \nodata  &  \nodata  & $29.15\pm 0.14$ &  \nodata  \\
\vspace{0.05cm}
$ 206.9$ &  \nodata  &  \nodata  & $29.67\pm 0.21$ &  \nodata  \\
\hline
\hline
\end{tabular}
\end{center}
\end{table*}

\begin{table*}
%
\caption{Surface brightness profiles of UGC02311}
%
\label{tab:UGC02311_sb}
\begin{center}
\renewcommand\tabcolsep{2.00000pt}
\begin{tabular}{ccccc}
\hline
\hline
Radius & $\mu_{g'}$ & $\mu_{r'}$ & $\mu_{i'}$ & $\mu_{z'}$ \\
$$[arcsec] & [mag arcsec$^{-2}$] & [mag arcsec$^{-2}$] & [mag arcsec$^{-2}$] & [mag arcsec$^{-2}$] \\
\hline
$   2.0$ & $19.95\pm 0.03$ & $19.23\pm 0.03$ & $18.83\pm 0.03$ & $18.57\pm 0.02$ \\
\vspace{0.05cm}
$   5.9$ & $20.95\pm 0.01$ & $20.21\pm 0.01$ & $19.80\pm 0.01$ & $19.57\pm 0.01$ \\
\vspace{0.05cm}
$  10.0$ & $21.59\pm 0.01$ & $20.92\pm 0.01$ & $20.55\pm 0.01$ & $20.34\pm 0.01$ \\
\vspace{0.05cm}
$  14.1$ & $21.87\pm 0.01$ & $21.32\pm 0.01$ & $20.96\pm 0.01$ & $20.81\pm 0.01$ \\
\vspace{0.05cm}
$  18.4$ & $22.47\pm 0.01$ & $21.91\pm 0.01$ & $21.57\pm 0.01$ & $21.42\pm 0.01$ \\
\vspace{0.05cm}
$  22.8$ & $23.15\pm 0.01$ & $22.53\pm 0.01$ & $22.13\pm 0.01$ & $21.99\pm 0.01$ \\
\vspace{0.05cm}
$  27.3$ & $23.51\pm 0.01$ & $22.86\pm 0.01$ & $22.46\pm 0.01$ & $22.31\pm 0.01$ \\
\vspace{0.05cm}
$  31.9$ & $23.82\pm 0.00$ & $23.21\pm 0.00$ & $22.81\pm 0.00$ & $22.68\pm 0.01$ \\
\vspace{0.05cm}
$  36.7$ & $24.10\pm 0.00$ & $23.57\pm 0.00$ & $23.21\pm 0.00$ & $23.11\pm 0.01$ \\
\vspace{0.05cm}
$  41.7$ & $24.53\pm 0.01$ & $24.04\pm 0.01$ & $23.70\pm 0.01$ & $23.63\pm 0.01$ \\
\vspace{0.05cm}
$  46.8$ & $25.10\pm 0.01$ & $24.61\pm 0.01$ & $24.24\pm 0.01$ & $24.18\pm 0.01$ \\
\vspace{0.05cm}
$  52.0$ & $25.78\pm 0.01$ & $25.23\pm 0.01$ & $24.86\pm 0.01$ & $24.83\pm 0.02$ \\
\vspace{0.05cm}
$  57.4$ & $26.59\pm 0.02$ & $26.05\pm 0.01$ & $25.63\pm 0.01$ & $25.66\pm 0.04$ \\
\vspace{0.05cm}
$  63.0$ & $27.32\pm 0.03$ & $26.74\pm 0.02$ & $26.42\pm 0.03$ & $26.52\pm 0.09$ \\
\vspace{0.05cm}
$  68.7$ & $27.82\pm 0.04$ & $27.32\pm 0.03$ & $26.81\pm 0.04$ & $27.16\pm 0.14$ \\
\vspace{0.05cm}
$  74.6$ & $28.32\pm 0.05$ & $27.63\pm 0.04$ & $27.50\pm 0.06$ & $27.77\pm 0.20$ \\
\vspace{0.05cm}
$  80.7$ & $28.67\pm 0.07$ & $27.90\pm 0.04$ & $27.71\pm 0.07$ &  \nodata  \\
\vspace{0.05cm}
$  87.0$ & $28.95\pm 0.08$ & $28.19\pm 0.05$ & $28.02\pm 0.08$ &  \nodata  \\
\vspace{0.05cm}
$  93.4$ & $28.62\pm 0.06$ & $28.08\pm 0.05$ & $27.87\pm 0.07$ &  \nodata  \\
\vspace{0.05cm}
$ 100.1$ & $29.62\pm 0.14$ & $28.76\pm 0.08$ & $29.25\pm 0.22$ &  \nodata  \\
\vspace{0.05cm}
$ 106.9$ & $29.37\pm 0.11$ & $28.87\pm 0.09$ &  \nodata  &  \nodata  \\
\vspace{0.05cm}
$ 113.9$ & $29.89\pm 0.16$ & $29.65\pm 0.16$ &  \nodata  &  \nodata  \\
\hline
\hline
\end{tabular}
\end{center}
\end{table*}

\end{document}